\documentclass[nofootinbib, twocolumn, longbibliography,superscriptaddress]{revtex4-2}

\usepackage{amsmath}
\usepackage{amsthm}
\usepackage{amssymb}
\usepackage{graphicx}
\usepackage[colorlinks=true, citecolor=blue]{hyperref}
\usepackage{color}
\usepackage{array}
\usepackage[makeroom]{cancel}
\usepackage[capitalize,nameinlink]{cleveref}
\usepackage{changes}
\usepackage{qcircuit}
\usepackage{braket}
\usepackage{bbm}
\usepackage{bm}
\usepackage{titlesec}
\usepackage{comment}
\usepackage{subfigure}

\newcommand{\beq}{\begin{equation}}
\newcommand{\eeq}{\end{equation}}
\newcommand{\PREP}{\mathrm{PREP}}
\newcommand{\SEL}{\mathrm{SEL}}

\renewcommand{\theparagraph}{\arabic{paragraph}.}

\titleformat{\paragraph}[runin]
  {\normalfont\itshape}
  {\theparagraph}
  {0.4em}
  {}
  
 \titlespacing{\paragraph}
  {10pt}{0.2\baselineskip}{0.3\baselineskip}

\begin{document}
\title{Simulating key properties of lithium-ion batteries with a fault-tolerant quantum computer}

\author{Alain Delgado}
\thanks{These authors contributed equally}
\affiliation{Xanadu, Toronto, ON, M5G 2C8, Canada}
\author{Pablo A. M. Casares}
\thanks{These authors contributed equally}
\affiliation{Departamento de F\'isica Te\'orica, Universidad Complutense de Madrid, Spain}
\author{Roberto dos Reis}
\affiliation{Xanadu, Toronto, ON, M5G 2C8, Canada}
\affiliation{Department of Materials Science and Engineering, Northwestern University, Evanston, IL 60208, United States}
\author{Modjtaba Shokrian Zini}
\affiliation{Xanadu, Toronto, ON, M5G 2C8, Canada}
\author{Roberto Campos}
\affiliation{Departamento de F\'isica Te\'orica, Universidad Complutense de Madrid, Spain}
\affiliation{Quasar Science Resources, SL}
\author{Norge Cruz-Hern\'andez}
\affiliation{Departamento de F\'isica Aplicada I, Escuela Polit\'ecnica Superior, Universidad de Sevilla, Seville, E-41011, Spain}
\author{Arne-Christian Voigt}
\affiliation{Volkswagen AG, Germany}
\author{Angus Lowe}
\affiliation{Xanadu, Toronto, ON, M5G 2C8, Canada}
\author{Soran Jahangiri}
\affiliation{Xanadu, Toronto, ON, M5G 2C8, Canada}
\author{M. A. Martin-Delgado}
\affiliation{Departamento de F\'isica Te\'orica, Universidad Complutense de Madrid, Spain}
\affiliation{CCS-Center for Computational Simulation, Universidad Politécnica de Madrid}
\author{Jonathan E. Mueller}
\affiliation{Volkswagen AG, Germany}
\author{Juan Miguel Arrazola}
\email{juanmiguel@xanadu.ai}
\affiliation{Xanadu, Toronto, ON, M5G 2C8, Canada}

\begin{abstract}
There is a pressing need to develop new rechargeable battery technologies that can offer higher energy storage, faster charging, and lower costs. Despite the success of existing methods for the simulation of battery materials, they can sometimes fall short of delivering accurate and reliable results. Quantum computing has been discussed as an avenue to overcome these issues, but only limited work has been done to outline how they may impact battery simulations. In this work, we provide a detailed answer to the following question: how can a quantum computer be used to simulate key properties of a lithium-ion battery? Based on recently-introduced first-quantization techniques, we lay out an end-to-end quantum algorithm for calculating equilibrium cell voltages, ionic mobility, and thermal stability. These can be obtained from ground-state energies of materials, which is the core calculation executed by the quantum computer using qubitization-based quantum phase estimation. The algorithm includes explicit methods for preparing approximate ground states of periodic materials in first quantization. We bring these insights together to perform the first estimation of the resources required to implement a quantum algorithm for simulating a realistic cathode material, dilithium iron silicate.
    
\end{abstract}
\maketitle

\section{Introduction}
\label{sec:intro}
Lithium-ion batteries have revolutionized portable electronic devices, allowing them to operate independently, safely, and over an extended period of time during multiple charging cycles~\cite{book_libs, yoshino2012birth,manthiram2017outlook,gur2018review,zubi2018lithium, kim2019lithium}.
Rechargeable batteries are also expected to play a central role in powering transportation and facilitating energy storage from renewable resources~\cite{2021electric_cars, fotouhi2016review, chen2009progress, trahey2020energy}. 
Despite their current remarkable performance, there is an increasing demand for new battery technologies that can deliver longer lifetimes, faster charging, higher capacity, and lower costs~\cite{choi2016promise, cheng2017toward, liu2017flexible, albertus2018status}. 

To achieve this goal, an interdisciplinary effort is crucial to discover new materials~\cite{de2014materials} and to understand their performance for batteries. Anchored by computer simulation methods, important steps have been taken towards reducing the overall cost associated to discovering and commercializing new materials. Electronic structure methods to simulate materials are widely employed to study the building blocks of commercial batteries: an electrochemical cell consisting of two electrodes, the anode and the cathode, separated by an electrolyte~\cite{grey2020prospects, 2021electric_cars, trahey2020energy}.

The ability to accurately compute ground-state energies of battery materials is important to derive key properties that define their performance. For example, accurate electronic structure calculations aid in the discovery of materials for high-energy cathodes~\cite{zhang2022pushing, cathode_materials_islam}, better anodes that enable faster charging~\cite{ahmed_fast_charging}, and more stable electrolytes~\cite{screening_electrolytes_jacs_2015}. Extending the lifespan of commercial batteries while maintaining their safe operation requires understanding how to suppress the reaction mechanisms driving the loss of ions and the degradation of the electrode active materials~\cite{leng2017hierarchical, xiao2019Li_dendrites}. For example, the growth of the solid electrolyte interphase consumes lithium ions which may result in a significant capacity loss~\cite{wang2018review, formation_sei}, and the electrochemical reduction of degraded oxide-based cathodes can lead to ignition of the electrolyte~\cite{thermal_runaway}. These simulations involve large-scale and costly computations to predict stable structures, phases, and properties of new materials~\cite{ceder2010opportunities, organics_electrodes}.

In materials science, density functional theory (DFT) methods offer an effective quantum mechanical description of the electronic structure of battery materials~\cite{hohenberg1964inhomogeneous,kohanoff2006electronic}. However, practical applications of DFT requires access to parametrized energy functionals of the electronic density which are only known approximately. This limits the ability of DFT to perform accurate simulations, especially for cathode materials with strong electronic correlations~\cite{urban2016computational}. Quantum computing is a fundamentally different approach to the simulation of quantum systems that may be capable of overcoming some of the limitations of DFT approximations~\cite{mcardle2020quantum, reiher2017elucidating, childs2018toward,ho2018promise}. Quantum algorithms are known to be capable of performing electronic structure calculations with chemical accuracy using time and memory resources that scale only polynomially with system size~\cite{berry2018improved, lee2021even, su2021fault}. Nevertheless, as with any emerging technology, quantum computing also faces several challenges. To fully unlock the potential of quantum computing, the long-term goal of the field is to build fault-tolerant devices capable of reliably implementing sophisticated large-scale quantum algorithms. This is a major experimental and theoretical effort requiring innovations on several fronts. While the computational resource requirements of quantum algorithms have steadily decreased over time~\cite{casares2021t}, there is still significant room to improve and identify problems of practical importance where a convincing argument can be made for the benefits of a quantum approach~\cite{reiher2017elucidating, von2021quantum, kim2021fault, goings2022reliably}.

In this work, we combine insights from quantum chemistry, materials science, and quantum algorithms to address the following question: how can a quantum computer be used to simulate key properties of a lithium-ion battery? Prior work at the intersection of battery simulation and quantum computing~\cite{kim2019lithium,rice2021quantum} focused mainly on computing ground-state energies of electrolyte molecules. We instead focus on the simulation of cathode materials, which is crucial for predicting important properties of a battery cell. We describe how the equilibrium cell voltage, ionic mobility, and thermal stability of lithium-ion batteries can be obtained from ground-state energy calculations of these materials. We then perform a detailed end-to-end description of a qubitization-based quantum phase estimation algorithm, which is based on the first-quantization techniques pioneered in Refs.~\cite{babbush2019quantum, su2021fault}. The description of the algorithm includes an explicit recipe for preparing approximate ground states in first quantization and a summary of circuit implementation strategies. We then apply the quantum algorithm to the concrete case study of dilithium iron silicate, a realistic cathode material. The analysis includes an estimate of the resources required to implement the full quantum algorithm, an estimate that is performed using the T-Fermion library~\cite{casares2021t}.

The rest of this manuscript is organized as follows. Due to the interdisciplinary nature of this work, we begin in Sec.~\ref{sec:background} with a comprehensive background on lithium-ion batteries, density functional theory, and quantum algorithms. We then shift gears in Sec.~\ref{sec:quantum_algo} and give a detailed end-to-end description of the full quantum algorithm to perform ground-state energy calculation of cathode materials. In Sec.~\ref{sec:application} we study how the quantum algorithm can be applied to the simulation dilithium iron silicate and report the results fo resource estimation. We conclude in Sec.~\ref{sec:conclusions}, followed by an outlook of future research directions in Sec.~\ref{sec:outlook}.

\begin{figure}[h!]
\centering
\includegraphics[width=1\columnwidth]{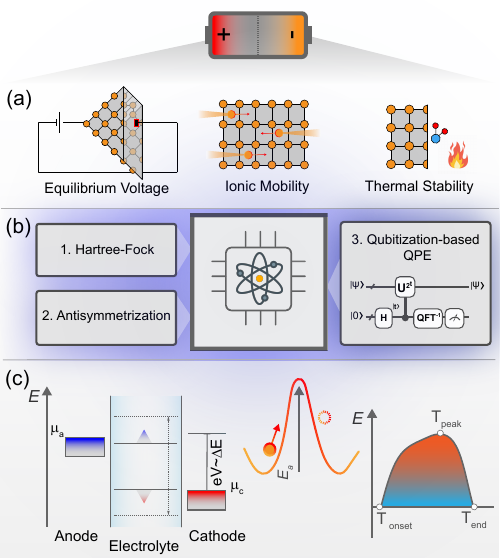}
\caption{\textbf{Quantum computing for battery simulations}. (a) Sketches depicting three key properties of lithium-ion batteries that can be obtained from calculations of the ground-state energies of cathode materials and isolated molecules (Sec.~\ref{sec:background}). (b) Summarizes the main steps of the first-quantized quantum algorithm implemented in this work. The ground-state energy $E$ of a given material is obtained by running a qubitization-based quantum phase estimation (QPE) algorithm on a quantum computer (Sec.~\ref{ssec:qubitization_qpe}). The initial state for the QPE method is obtained by calculating Hartree-Fock orbitals and using the quantum computer to prepare the corresponding anti-symmetric Hartree-Fock state (Sec.~\ref{ssec:antisym}). (c) Shows examples of measurable quantities that can be derived: The cell voltage is given by the difference between the chemical potentials ($\mu$) of the electrodes computed from the energy variation ($\Delta E$) of the cathode material; the activation energy ($E_a$), which is used to predict the ionic mobility; and the temperature profile that helps to define the battery thermal stability.}
\label{fig:overview-schema}
\end{figure}

\section{Background}
\label{sec:background}

This work encompasses technical information from several disciplines: materials science, computational chemistry, battery technologies, and quantum algorithms. Knowledge from all these fields is important to understand how quantum computing can be used in the context of battery simulations. While experts might choose to skip some of these sections, it contains important information that is widely used throughout this work.

\subsection{Lithium-ion batteries}
\label{ssec:libs}
\begin{figure}[t]
\centering
\includegraphics[width=1\columnwidth]{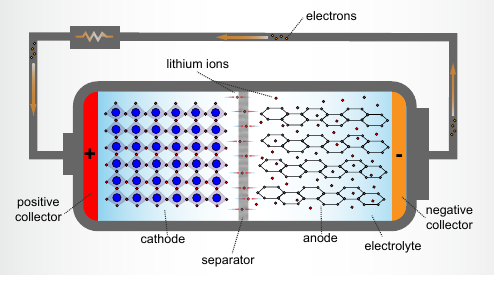}
\caption{\textbf{Schematic of a typical lithium-ion battery}. The negative electrode is usually a graphitic carbon that holds lithium ions within its layers, whereas the positive electrode is a source of lithium ions. During discharging, lithium ions move within the battery from the negative to the positive electrode. The process is reversed during charging. The electrolyte transports lithium ions between the electrodes and the separator functions as a physical barrier keeping cathode and anode apart. Negative and positive collectors receive electrons from the external circuit during charging and discharging, respectively.}
\label{fig:lib-schema}
\end{figure}

Fig.~\ref{fig:lib-schema} depicts the fundamental components of a rechargeable lithium-ion battery~\cite{goodenough2018we}. The battery cell consists of a positive electrode (cathode) and a negative electrode (anode) that are electrically isolated by a porous membrane (separator) and embedded in an ion-conducting material (electrolyte). The conversion of chemical into electrical energy in a battery cell is driven by the chemical reactions that occur at the electrode-electrolyte interface. During discharge, an oxidation reaction at the anode produces electrons and lithium ions. The electrons flow via an external circuit and the lithium ions diffuse through the electrolyte until they get inserted into the cathode material (intercalation), which is reduced by the external electrons. During charging, an external voltage is applied to reverse this process, i.e., the lithium ions are extracted from the cathode (deintercalation), transported in the opposite direction, and intercalated into the anode material.

Typically, the cathode contains active materials based on metal-oxides. There main classes of cathode materials are layered and spinel oxides such as lithium cobalt ($\mathrm{LiCoO}_2$) and lithium manganese ($\mathrm{LiMn}_2\mathrm{O}_4$) oxide cathodes, and the polyanion materials, e.g., $\mathrm{Li}_2\mathrm{FeSiO}_4$~\cite{manthiram2020reflection}. Their chemical composition and main distinguishing features are discussed in more details in Sec.~\ref{sec:application}. The commercial active materials of the anode are typically carbon-based materials, e.g., graphite and amorphous carbon, as they offer a safe, environmentally friendly, and cost-efficient option. However, carbon-based anodes possess a low specific capacity that lowers the overall capacity of the battery . Alternatively, silicon, germanium, and tin have also been actively investigated as high-capacity anode materials~\cite{nitta2014high}. The electrolyte, whose main role is to efficiently transport lithium ions between the electrodes, typically consists of a lithium salt such as lithium hexafluorophosphate ($\mathrm{LiPF}_6$) dissolved in high dielectric solvents like ethylene carbonate~\cite{wang2015development}. The separator serves as a physical barrier keeping the cathode and anode apart, preventing the direct flow of electrons, and allowing only the lithium ions to pass through. Commercial separators are typically synthetic resin such as polyethylene (PE) and polypropylene (PP). Typically, the anode materials offer a higher lithium-ion storage capacity than cathodes. Therefore, the cathode material is the main limiting factor in the performance of batteries~\cite{tarascon2011issues} and also responsible for up to 50$\%$ of the total battery cost~\cite{li2020high}.

Optimization of lithium-ion battery performance is critical to develop the next generation of energy storage systems. Such advances depend not only on the discovery of novel materials but also on the development of more accurate methods to simulate key properties of lithium-ion batteries. The landscape of properties determining their performance is extremely rich. It includes mechanical and electrochemical properties, thermal stability of the cathode, the electrochemical windows of the electrolyte, formation of the solid-electrolyte interphase, and ionic mobility, among others~\cite{lee2013roles, kermani2017characterization, yu2018electrode, van2001lithium}. Typically, the computational simulation of these properties requires multi-scale approaches combining electronic structure methods, molecular dynamics, and continuum models to describe solvation effects~\cite{urban2016computational, wang2018review, manthiram2020reflection}. In this section, we follow the strategy presented in Ref.~\cite{urban2016computational} and focus our attention on the equilibrium cell voltage of the battery, the ionic mobility, and the thermal stability of the cathode material. We briefly describe these properties and explain how they can be computed from the ground-state energy of the cathode material. This section is thus a summary of the main results of Ref.~\cite{urban2016computational}, where readers can find a more in-depth discussion of these properties.

\subsubsection{The equilibrium voltage}

The equilibrium voltage is key to determining the amount of energy that can be stored in a battery in comparison to its volume (energy density) and weight (specific energy)~\cite{voltage_energy_density}. The average voltage $V$ of a device that produces electrical energy from chemical reactions (electrochemical cell) is given by the Nernst equation \cite{feiner1994nernst, barnard1980studies}
\begin{equation}
V = -\frac{\Delta G}{nF},
\label{eq:nernst}
\end{equation}
where $n$ is the number of charges transferred, $F$ is the Faraday constant, and $\Delta G$ is the variation of the free energy associated with the cell reaction. For example, for the typical lithium cobalt oxide $\mathrm{LiCoO}_2$ cathode and a metallic lithium anode, the electrical work generated by the chemical reaction
\begin{equation}
\mathrm{Li}_{x_1}\mathrm{Co}\mathrm{O}_2 + (x_2-x_1)\mathrm{Li} \rightarrow \mathrm{Li}_{x_2}\mathrm{Co}\mathrm{O}_2,
\label{eq:cell_reaction}
\end{equation}
is determined by the free energy difference
\begin{equation}
\Delta G = G_{\mathrm{Li}_{x_2}\mathrm{Co}\mathrm{O}_2} - G_{\mathrm{Li}_{x_1}\mathrm{Co}\mathrm{O}_2} - (x_2-x_1) G_\mathrm{Li},
\label{eq:free_energy}
\end{equation}
where $x_2 > x_1$ denotes the number of ions per formula unit in the cathode material upon lithium insertion. At low temperatures ($\leq 300K$), the thermal and entropic contributions to the free energy are small~\cite{urban2016computational, ma2018computer}, and the equilibrium cell voltage is computed in terms of the variation of the internal energy $\Delta E$. In practice, the average cell voltage is estimated by taking the energy difference for the extreme cases in which the amount of intercalated lithium ions in the material's unit cell is maximum (lithiated phase) and minimum (delithiated phase). For example, predicting the voltage for $\mathrm{LiCoO}_2$ requires calculating the total energies of two materials with compositions $\mathrm{LiCoO}_2$ and $\mathrm{CoO}_2$. This leads to the following expression for the equilibrium voltage
\begin{equation}
V = - \frac{ \left[ E_{\mathrm{LiCoO}_2} - E_{\mathrm{CoO}_2} - E_\mathrm{Li} \right]}{F}.
\label{eq:voltage_DE}
\end{equation}
The total energies entering Eq.~\eqref{eq:voltage_DE} are usually obtained from electronic structure calculations performed using density functional theory (DFT) methods. Current limitations of DFT to predict accurate voltages are discussed in more detail in Sec.~\ref{sec:dft}.

\subsubsection{Ionic Mobility}
Developing high-power batteries requires using materials that allow for an optimal and stable mobility of the lithium ions during battery operation~\cite{braun2012high, alikin2018quantitative, saito2019factors}. To this aim, understanding the microscopic mechanisms that determine the ionic mobility in the electrode materials is essential for predicting new materials with better lithium intercalation rates which can enable, for example, faster charging regimes~\cite{urban2016computational}.

The relevant quantity that characterizes the mobility of the lithium ions in a given material is the chemical diffusivity $D$. In cases where the diffusion mechanisms do not depend on the temperature, a microscopic model can be used to describe the hopping of a lithium ion from its original site to a neighboring vacant site in the crystal structure of the host material. In this approximation, the diffusivity can be calculated as~\cite{vineyard1957frequency}
\begin{equation}
    D(T)\approx a^2 \nu^* e^{-\frac{(E_\mathrm{T} - E_\mathrm{I})}{k_B T}}, 
\label{eq:diffusivity}
\end{equation}
where $a$ is the hopping distance between two adjacent sites~\cite{kutner1981chemical}, $\nu^*$ is the average vibration frequency of the lithium ions in the material (effective attempt frequency)~\cite{de2018analysis, van2008nondilute}, $E_\mathrm{I}$ is the total energy of the material when the lithium ion is in the original site, $E_\mathrm{T}$ is the energy of the transition state that has to be overcome during the diffusion, and $T$ and $k_\mathrm{B}$ are the temperature and the Boltzmann constant, respectively. For given initial and final states of the hopping process, there are efficient methods such as the nudged elastic band approach~\cite{neb_method} to find the transition state along the minimum energy path of the diffusion process. From Eq.~\eqref{eq:diffusivity}, it suffices to compute the activation energy $(E_T - E_I)$ to predict the ionic diffusivity.

\subsubsection{Thermal stability of cathode materials}
There are many different processes that contribute to the degradation of the battery performance over time. These include the formation of the solid electrolyte interphase, degradation of the cathode active materials, lithium plating on the anode, and growth of lithium dendrites, among others~\cite{leng2017hierarchical}. Simulating these processes remains a challenge since it involves bottom-up approaches from the atomic level to the macroscopic scale~\cite{hausbrand2015fundamental}.

Predicting the thermal stability of the cathode materials is important to maximize the safety of lithium-ion batteries, which can be unstable in their charged state. As more lithium ions are removed from oxide-based cathode materials, they may degrade to other phases of the material~\cite{wang2007first}. This phase transformation, typically driven by an exothermic chemical reaction, can result in the release of heat and oxygen gas, which in turn may lead to thermal runaway and combustion of the electrolyte~\cite{thermal_runaway, zheng2015structural}.

The reduction chemical reaction for a lithium metal oxide cathode with composition $\mathrm{Li}_x\mathrm{M}_y\mathrm{O}_{z+z^\prime}$, where M refers to one or multiple transition metals and O refers to oxygen, is given by~\cite{urban2016computational} 
\begin{equation}
\mathrm{Li}_x\mathrm{M}_y\mathrm{O}_{z+z^\prime} \rightarrow  \mathrm{Li}_x\mathrm{M}_y\mathrm{O}_z \text{+} \frac{z^\prime}{2}\mathrm{O}_2.
\label{eq:red_reaction}
\end{equation}
The free energy change of the reaction in Eq.~\eqref{eq:red_reaction} is given by
\begin{equation}
\Delta G = -G_{\mathrm{Li}_x\mathrm{M}_y\mathrm{O}_{z+z^\prime}} + G_{\mathrm{Li}_x\mathrm{M}_y\mathrm{O}_z} + \frac{z^\prime}{2} G_{\mathrm{O}_2}.
\label{eq:free_thermal_stability}
\end{equation}
Under isobaric and isothermal conditions, the free energy difference $\Delta G$ can be written as
\begin{equation}
\Delta G = \Delta E + P \Delta V - T \Delta S,
\label{eq:deltag}
\end{equation}%
where $P$ is the pressure, $\Delta V$ is the change in the volume of the material, $T$ is the temperature, and $\Delta S$ is the change in entropy. The dominant contributions to Eq.~\eqref{eq:deltag} come from the variation of the internal energy $\Delta E$ of the cathode material and the entropy change due to the release of oxygen gas \cite{wang2007first}. Thus, the reaction free energy can be approximated as

\begin{eqnarray}
\Delta G \approx -E_{\mathrm{Li}_x\mathrm{M}_y\mathrm{O}_{z+z^\prime}} + E_{\mathrm{Li}_x\mathrm{M}_y\mathrm{O}_z} \nonumber \\
+ \frac{z^\prime}{2} E_{\mathrm{O}_2} -\frac{z^\prime}{2} T S(\mathrm{O}_2),
\label{eq:free_approx}
\end{eqnarray}
where $E_{\mathrm{O}_2}$ is the total energy of the oxygen molecule at zero temperature and $S(\mathrm{O}_2)$ is its entropy, which can be obtained from experimental thermochemistry data~\cite{thermo_tables_1998}. The temperature for which $\Delta G$ equals zero is the temperature at which the cathode material becomes unstable and undergoes the degradation reaction in Eq.~\eqref{eq:red_reaction}. This temperature can be calculated from the equation above as:

\begin{equation}
T = \frac{-E_{\mathrm{Li}_x\mathrm{M}_y\mathrm{O}_{z+z^\prime}} + E_{\mathrm{Li}_x\mathrm{M}_y\mathrm{O}_z} + (z^\prime/2) E_{\mathrm{O}_2} }{(z^\prime/2) S({\mathrm{O}_2})}.
\label{eq:temperature}
\end{equation}
Similar to Eq.~\eqref{eq:voltage_DE} for computing the equilibrium cell voltage, Eq.~\eqref{eq:temperature} allows us to compute the transition temperature for a given cathode material by calculating the ground state energies of different phases of the cathode material and the oxygen molecule. A key step for assessing the thermal stability of a cathode material is the construction of a phase diagram to reliably predict the stability of the reduced phases of the material~\cite{wang2007first}. This requires highly accurate calculations of formation energies which is challenging for current DFT methods~\cite{wang2006oxidation}.

\subsection{Density functional theory}
\label{sec:dft}
At present, first-principles calculations of the electronic structure of cathode materials are largely performed using density functional theory (DFT) methods~\cite{urban2016computational, cathode_materials_islam}. In this section we explain the main concepts of DFT and describe the most common approximations used for battery simulations.

\subsubsection{Basic concepts}
\label{ssec:basics_dft}
Density functional theory has been the workhorse for simulating the electronic structure of molecules and materials for more than two decades~\cite{burke_perspective_DFT}. The core of this success is that the quantity used in DFT to compute the properties of an interacting electron system is the ground-state electronic density $n(\bm{r})$ (a function of three variables), which is a much simpler object than the wave function $\Psi(\bm{r}_1, \dots ,\bm{r}_\eta)$ (a function of $3\eta$ variables). The ground-state wave function $\Psi_0$ of the $\eta$-electron system is a solution of the Schr\"odinger equation
\begin{equation}
H \Psi_0(\bm{r}_1, \dots ,\bm{r}_\eta) = E_0 \Psi_0(\bm{r}_1, \dots , \bm{r}_\eta),
\label{eq:schrodinger}
\end{equation}
where $E_0$ is the ground-state energy and $H = T + U + V$ is the electronic Hamiltonian. Here $T$ is the kinetic-energy operator, $U$ is a given potential operator, e.g., the electron-nuclei interaction, and $V$ is the Coulomb electron-electron interaction. These terms are defined as follows:
\begin{eqnarray}
&& T = \sum_{i=1}^\eta -\frac{\nabla_i^2}{2}, \label{eq:dft_T} \\
&& U = \sum_{i=1}^\eta u(\bm{r}_i), \\
&& V = \frac{1}{2}\sum_{i\neq j=1}^\eta \frac{1}{\vert\vert \bm{r}_i - \bm{r}_j \vert\vert} \label{eq:dft_V}.
\label{eq:operators}
\end{eqnarray}

The electronic structure problem defined above can be recast in terms of the electronic density $n(\bm{r})$, which is given by
\begin{equation}
n(\bm{r}) = \eta \int d\bm{r}_2 \dots d\bm{r}_\eta \vert\vert \Psi(\bm{r}, \bm{r}_2, \dots, \bm{r}_\eta) \vert\vert^2.
\label{eq:density}
\end{equation}
The Hohenberg-Kohn theorem~\cite{hohenberg1964inhomogeneous}, one of the pillars of DFT, proves that there is a one-to-one correspondence between the potential $u(\bm{r})$ and the ground-state electronic density $n_0(\bm{r})$ of the interacting system. This mapping is expressed by writing $u(\bm{r})$ as a density functional $u[n_0](\bm{r})$, which implies that the wave function $\Psi_0$ computed via the Schr\"odinger Eq.~\eqref{eq:schrodinger} also becomes a density functional $\Psi_0[n_0](\bm{r})$. Thus, the expectation value of the Hamiltonian for a given potential $u_0$ defines the energy functional

\begin{eqnarray}
E_{u_0}[n] &=& \langle \Psi[n] \vert T + U_0 + V \vert \Psi[n] \rangle \nonumber\\
&=& \int d\bm{r}~n(\bm{r}) u_0(\bm{r}) + F[n],
\label{eq:energy_func}
\end{eqnarray}
where $\Psi[n]$ is a wave function producing the density $n(\bm{r})$ and $F[n]=T[n] + V[n]$ is the so-called universal functional, since it is the same for any Coulomb system. Applying the variational principle, the ground-state electronic density $n_0(\bm{r})$ can be obtained by solving the Euler equation

\begin{equation}
\frac{\delta}{\delta n(\bm{r})} \left[ E_{u_0}[n] -\mu \int d\bm{r}^\prime n(\bm{r}^\prime) \right] = 0,
\label{eq:euler}
\end{equation}
where $\mu$ is a Lagrange multiplier to ensure the correct number of electrons. In principle, Eq.~\eqref{eq:euler} allows us to find the electronic density $n_0(\bm{r})$ and thus the ground-state energy $E_{u_0}[n_0]$ without having to solve the Schr\"odinger equation. However, the actual form of the universal functional $F[n]$ is unknown and must be approximated.

It was the later formulation proposed by Kohn and Sham~\cite{kohn_sham_dft} which transformed DFT into a practical computational scheme for simulations. The Kohn-Sham approach assumes that for any interacting system with ground-state density $n(\bm{r})$ there is always a non-interacting system, the Kohn-Sham system, which reproduces the same density $n(\bm{r})$. Since the Hamiltonian of the non-interacting system contains no $V$ operator, Eq.~\eqref{eq:energy_func} for the Kohn-Sham system simplifies to

\begin{equation}
E_{u_s}[n] = \int d\bm{r} \; n(\bm{r}) u_s(\bm{r}) + T_s[n],
\label{eq:func_ks}    
\end{equation}
where $u_s(\bm{r})$ is an effective potential and $T_s[n]$ is the kinetic energy functional for the non-interacting system. In this case, the density $n(\bm{r})$ can be found from the equation
\begin{equation}
\frac{\delta E_{u_s}[n]}{\delta n(\bm{r})} = \frac{\delta T_s[n]}{\delta n(\bm{ r})} + u_s(\bm{r}) = \mu.
\label{eq:euler_ks}
\end{equation}
Solving the equation above still requires access to the functional $T_s[n]$, which is only approximately known~\cite{gross2013density}. However, the many-body wave function for the non-interacting system is the product sate $\phi_1(\bm{r}_1) \phi_2(\bm{r}_2) \dots \phi_\eta(\bm{r}_\eta)$, antisymmetrized under all possible particle exchanges (a Slater determinant~\cite{jensen2017introduction}), where the single-electron states $\phi_i(\bm{r})$ satisfy the equation
\begin{equation}
\left[ -\frac{1}{2} \nabla^2 + u_s(\bm{r}) \right] \phi_i(\bm{r}) = \epsilon_i \phi_i(\bm{r}),
\label{eq:ks_equations}
\end{equation}
where $\epsilon_i$ is the energy of the Kohn-Sham orbital $\phi_i(\bm{r})$, and the electronic density $n(\bm{r})$ can be computed as
\begin{equation}
n(\bm{r}) = \sum_{i=1}^\eta \vert\vert \phi_i(\bm{r}) \vert\vert^2.
\label{eq:ks_density}
\end{equation}
In order to solve Eq.~\eqref{eq:ks_equations} we need to know the effective potential $u_s(\bm{r})$. To that aim, we first rewrite the energy functional $E_{u_0}[n]$ in Eq.~\eqref{eq:energy_func}, as
\begin{equation}
\kern-5pt E_{u_0}[n] = T_s[n] + \int d\bm{r} n(\bm{r}) u_0(\bm{r}) + E_\mathrm{H}[n] + E_\mathrm{xc}[n],
\label{eq:func_xc}
\end{equation}
where
\begin{equation}
E_\mathrm{H}[n] = \frac{1}{2} \int d\bm{r}\;d\bm{r}^\prime \frac{n(\bm{r}) n(\bm{r}^\prime)}{\vert\vert \bm{r} - \bm{r}^\prime \vert\vert},
\label{eq:hartree}
\end{equation}
is the classical Coulomb energy, and $E_\mathrm{xc}[n]$ is called the {\it exchange-correlation} (xc) energy functional defined as
\begin{equation}
E_\mathrm{xc}[n] = T[n] - T_s[n] + V[n] - E_\mathrm{H}[n].
\label{eq:def_xc}
\end{equation}
Then, by taking the derivative of the functional $E_{u_0}[n]$ in Eq.~\eqref{eq:euler} and comparing the result with Eq.~\eqref{eq:euler_ks}, we find that the Kohn-Sham potential $u_s(\bm{r})$ is given by
\begin{equation}
u_s[n](\bm{r}) = u_0(\bm{r}) + u_\mathrm{H}[n](\bm{r}) + u_\mathrm{xc}[n](\bm{r}),
\label{eq:ks_pot}
\end{equation}
where
\begin{equation}
\kern-5pt u_\mathrm{H}[n](\bm{r}) = \int d\bm{r}^\prime \frac{n(\bm{r}^\prime)}{\vert\vert \bm{r} - \bm{r}^\prime \vert\vert} ~~ \mathrm{and} ~~ u_\mathrm{xc}[n](\bm{r})=\frac{\delta E_\mathrm{xc}[n]}{\delta n(\bm{r})},
\label{eq:v_xc}
\end{equation}
are the Hartree and exchange-correlation potentials, respectively. Eqs.~\eqref{eq:ks_equations}, \eqref{eq:ks_density} and \eqref{eq:ks_pot} are known as the Kohn-Sham equations, and the solutions of Eq.~\eqref{eq:ks_equations} (Kohn-Sham orbitals) for the exact exchange-correlation functional can be used to compute the ground-state electronic density $n_0(\bm{r})$ of the interacting quantum system and its total energy $E_{u_0}[n_0]$ given by \cite{ullrich2011time}
\begin{eqnarray}
E_{u_0}[n_0] = && \sum_{i=1}^\eta \epsilon_i - E_\mathrm{H}[n_0] \nonumber\\ && - \int d\bm{r} n_0(\bm{r}) u_\mathrm{xc}[n_0](\bm{r}) + E_\mathrm{xc}[n_0].
\label{eq:dft_energy}
\end{eqnarray}

\subsubsection{DFT approximations for simulating cathode materials}
\label{ssec:dft_approx}
In practice, we only have access to approximate exchange-correlation energy functionals. Different strategies included in the so-called Jacob's ladder to DFT approximations~\cite{perdew2001jacob} have been used to develop a diverse landscape of available functionals~\cite{burke_perspective_DFT, scuseria2005progress}. In general, choosing the best possible approximation to simulate a particular system is never straightforward and typically requires expert knowledge for making the most adequate choices. In particular, most DFT simulations of the battery properties described in the previous section rely on the local-density (LDA) and generalized-gradient (GGA) approximations~\cite{urban2016computational}. Within the LDA approach, the exchange and correlation energies are computed from the value of the electronic density at each point using analytical expressions derived for the uniform electron gas~\cite{ueg}. On the other hand, GGA functionals accounts for the inhomogeneities of the electronic structure of materials by including terms that depend on the gradient of the density.

DFT has been central to make progress in the atomistic simulation of materials even though it faces important limitations in simulating key properties of batteries. LDA and GGA functionals are extensively used in materials science due to their favorable computational performance. These approximations are affected by the well known self-interaction error \cite{sie_dft} which can lead to very inaccurate values of the absolute energy of the quantum system. However, most properties of molecules and materials are obtained by computing total energy differences where these errors cancel as long as the electronic structure of the system does not change drastically. This is not the case when simulating the battery properties described in the previous section. 

Lithium insertion into the cathode material involves an electron transfer from a metallic state in a lithium anode to a localized state in the transition metal-oxide cathode. Using DFT to compute the energy difference between these two different electronic phases lacks error cancellation and leads to large deviations of up to one volt in the calculated voltages~\cite{urban2016computational}. For the same reasons, it is still challenging for DFT to compute accurate formation energies for reliably predicting other stable phases of the cathode materials.

The DFT+U method~\cite{anisimov1997dftU} has been used to partially mitigate this problem of standard DFT approximations. Inspired by the Hubbard model, this approach incorporates a Hubbard-like term to treat the strong on-site Coulomb interactions between the electrons populating the $d$ or $f$ orbitals~\cite{anisimov1997dftU}. However, the improvements on the simulated quantities comes at the price of using specific values of the Hubbard parameter $U$ which are strongly system-dependent. Alternatively, LDA/GGA self-interaction error can also be reduced by using specific hybrid functionals which incorporate a fraction of the exact exchange from Hartree-Fock theory~\cite{heyd2003hybrid}. However, hybrid functionals also contain an adjustable parameter to select the amount of exact exchange to be included in the calculation, and their computational performance scales poorly with the system size. Overall, these corrections to DFT approximations reduce the deviations in the predicted cell voltages to about $0.2$ volts \cite{chevrier2010hybrid, urban2016computational}.

\subsection{Quantum algorithms for ground-state energies}
\label{ssec:qalg}
Since their introduction, quantum computers have been proposed to tackle problems related to the simulation of quantum systems that are intractable for classical methods~\cite{nielsen2002quantum}. In a nutshell, the goal of quantum algorithms is to
directly prepare wave functions describing molecules or materials and then proceed to extract information from them, such as the expectation values of relevant observables. This is achieved by representing states of fermionic systems as states of qubits on a quantum computer, an approach that is memory efficient because the number of required qubits scales linearly with the number of particles in the system~\cite{mcardle2020quantum}. The challenge for quantum algorithms is to identify concrete and efficient methods for wave function preparation and information extraction. Here we focus on methods that can be used to compute the ground-state energies of cathode materials, the key quantity to simulate the battery properties described in Sec.~\ref{ssec:libs}.

Some of the main quantum algorithms for estimating ground-state energies are the variational quantum eigensolver~\cite{peruzzo2014variational, kandala2017hardware}, adiabatic algorithms~\cite{aspuru2005simulated, albash2018adiabatic}, imaginary-time evolution~\cite{motta2020determining, mcardle2019variational}, quantum Metropolis sampling~\cite{temme2011quantum}, Krylov subspace methods~\cite{cortes2021quantum}, and quantum phase estimation~\cite{abrams1999quantum, nielsen2002quantum, reiher2017elucidating, berry2018improved}. In this work, we focus on the quantum phase estimation algorithm. As we explain in more detail shortly, provided with an approximate description of the ground state, quantum phase estimation can calculate ground-state energies with complexity inversely proportional to the error and with overall cost scaling polynomially with system size -- a performance that no classical algorithm is known to be able to replicate. The main steps of this method are described below.

\subsubsection{Quantum Phase Estimation}\label{ssec:QPE}

We assume that we are given (i) a quantum circuit that can implement a unitary operator $U$, and (ii) an input eigenstate $|\psi\rangle$ of $U$ such that $U|\psi\rangle=e^{i\theta}|\psi\rangle$. The goal is to estimate the phase $\theta$ with precision $\varepsilon$. If we can solve this phase estimation problem, then it is also possible to estimate eigenvalues $E_k$ of an electronic Hamiltonian $H$. This is done by choosing $U$ to share eigenvectors with $H$ such that the eigenphases $\theta_k$ can be related to the eigenvalues of $H$ by an invertible function $\theta_k=f(E_k)$. For example, a simple strategy to accomplish this is to set $U=e^{-iH}$.

\begin{figure}
\mbox{
\Qcircuit @C=0.9em @R=0.75em {
  & \qw & \multigate{3}{U} & \multigate{3}{U^2} & \qw & \cdots & & \multigate{3}{U^{2^t}} & \qw \\
\lstick{\ket{\psi}} \gategroup{1}{1}{3}{1}{.5em}{\{}& \qw & \ghost{U} & \ghost{U^2} & \qw & \cdots & & \ghost{U^{2^t}} &  \rstick{\ket{\psi}} \qw \gategroup{1}{9}{3}{9}{.5em}{\}} \\
 & \qw & \ghost{U} & \ghost{U^2} & \qw & \cdots & & \ghost{U^{2^t}} &  \qw \\
& & & & & & & & & & &\\
& & & & & & & & & & &\\
& & & & & & & & & & &\\
\lstick{\ket{0}} & \gate{\text{H}} & \ctrl{-3} & \qw & \qw & \cdots & & \qw & \multigate{4}{QFT^{-1}} & \qw & \meter\\
\lstick{\ket{0}} & \gate{\text{H}} & \qw & \ctrl{-4} & \qw & \cdots & & \qw & \ghost{QFT^{-1}} & \qw & \meter\\
& \vdots & & & & \ddots & & & & & \vdots\\
& & & & & & & & & & \\
\lstick{\ket{0}} & \gate{\text{H}} & \qw & \qw & \qw & \cdots & & \ctrl{-7} & \ghost{QFT^{-1}} & \qw & \meter
 \\
& & & & & & & & & & & \\
}
}
\caption{\textbf{Conceptual circuit diagram for the quantum phase estimation algorithm}. The system register is initialized in the eigenstate $\ket{\psi}$ of the target unitary. The auxiliary register consists of $t$ qubits, initialized in the basis state $\ket{0}$. After applying a Hadamard gate (H) on all auxiliary qubits, increasing powers of the target unitary are applied to the system register, controlled on the state of each auxiliary qubit. Concluding with an inverse quantum Fourier transform (QFT$^{-1}$) and measuring the output qubits gives a binary string that can be processed to estimate the desired phase. }
\label{fig:qpe}
\end{figure}
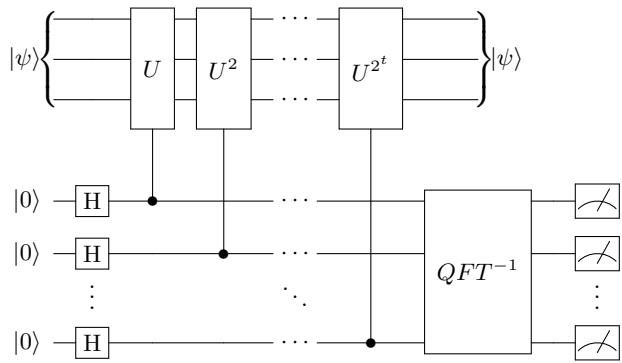

The standard version of the quantum phase estimation algorithm uses two registers, as sketched in Fig.~\ref{fig:qpe}. The first register contains the qubits required to represent the state $\ket{\psi}$, and the second register contains $t$ auxiliary qubits. The main strategy of the algorithm is to prepare the state
\beq\label{eq:qpe_state}
\ket{\Psi}=\frac{1}{\sqrt{2^t}}\left(\sum_{k=0}^{2^t-1}e^{2\pi i \phi k} \ket{k}\right)\ket{\psi}, 
\eeq
where $\theta=2\pi \phi$ , $0\leq\phi\leq 1$, and $\ket{k}$ denotes a computational basis state for $t$ auxiliary qubits. In Eq.~\eqref{eq:qpe_state}, the state of the auxiliary qubits is equivalent to applying a quantum Fourier transform~\cite{nielsen2002quantum} to a state $\ket{\mathrm{bin}(\phi)}=\ket{\phi_{1}, \phi_{2}, \dots, \phi_{t}}$ that encodes a binary representation of $\phi$ as $\phi=\sum_{j=1}^t\phi_j 2^{-j}$, where $\phi_j\in\{0,1\}$.
Therefore, applying an inverse quantum Fourier transform to the auxiliary qubits of the state $\ket{\Psi}$ in Eq.~\eqref{eq:qpe_state} will prepare the state $\ket{\mathrm{bin}(\phi)}$:
\beq
\frac{1}{\sqrt{2^t}}\sum_{k=0}^{2^t-1}e^{2\pi i \phi k} \ket{k}\ket{\psi} \xrightarrow{QFT^{-1}} \ket{\mathrm{bin}(\phi)}\ket{\psi}.
\eeq

The auxiliary qubits can then be measured in the computational basis to retrieve $\phi$ and thus the phase $\theta=2\pi i \phi$. The number of auxiliary qubits required to approximate $\phi$ to an accuracy $\varepsilon=2^{-n}$ with probability of success $1-\delta$ is at most $t = n + \left \lceil{\log \left( 2+\frac{1}{2\delta} \right)} \right \rceil$~\cite{nielsen2002quantum}. Here the logarithm is in base 2, a choice maintained throughout this article.

The task of phase estimation can therefore be reduced to preparing the state $\ket{\Psi}$ in Eq.~\eqref{eq:qpe_state}. A quantum computer can prepare this state as follows. First, a Hadamard gate is applied to all auxiliary qubits to create an equal superposition over the computational basis states:
\beq
\ket{0}\ket{\psi_k}\rightarrow \frac{1}{\sqrt{2^t}}\sum_{k=0}^{2^t-1} \ket{k}\ket{\psi}.
\eeq

As shown in Fig.~\ref{fig:qpe}, we then apply $U$ controlled on the state of the first auxiliary qubit, $U^2$ controlled on the state of the second qubit, $U^4$ controlled on the third qubit, and so forth until applying $U^{2^t}$ controlled on the final qubit. Denoting the states of the auxiliary qubits as $\ket{k}=\ket{k_0,k_1\ldots, k_t}$ such that the bit string $k_0k_1\ldots k_t$ is a binary representation of the integer $k$, this sequence of operations transforms the state of both registers as
\begin{align}
\frac{1}{\sqrt{2^t}}\sum_{k=0}^{2^t-1} \ket{k}\ket{\psi}\rightarrow&\frac{1}{\sqrt{2^t}}\sum_{k=0}^{2^t-1} e^{i \theta (\sum_{j=1}^t k_j2^j)}\ket{k}\ket{\psi}\nonumber\\
=&\frac{1}{\sqrt{2^t}}\sum_{k=0}^{2^t-1} e^{2\pi i \phi k}\ket{k}\ket{\psi},
\end{align}
as desired.

To perform simulations of materials, a system initialization into the ground state of the Hamiltonian would be required to carry out the algorithm as depicted above. This is of course generally not feasible in practice. Instead, consider a general input state $\ket{\psi}=\sum_{i}c_i\ket{E_i}$ expressed in the eigenbasis of the encoded Hamiltonian, where the eigenstate $\ket{E_i}$ has a corresponding eigenvalue $E_i$. After encoding the Hamiltonian into a suitable unitary, the algorithm produces an estimate of $E_i$ with probability $p_i=|c_i|^2$. In particular, it provides an estimate of the ground-state energy with probability $p_0=|c_0|^2=|\braket{\psi|E_0}|^2$. 
Thus, it is extremely important that the input state has a sufficiently large overlap with the ground state since, on average, the quantum phase estimation algorithm needs to be repeated $O(1/p_0)$ times to calculate the ground-state energy with high probability. It is often possible to use classical methods to compute an upper bound on the ground-state energy which is smaller than the first excited-state energy. Hence, the accuracy of the phase estimation procedure can be gradually increased, and as soon as the energy is estimated to be above the upper bound with high probability, the algorithm is restarted~\cite{berry2018improved}. This alleviates the cost of running the algorithm when failing to project onto the ground state.

Applying the controlled unitaries is the most expensive part of the algorithm because they can be complicated operations that depend on all parameters of the electronic Hamiltonian. Nevertheless, as we discuss in more detail in Sec.~\ref{sec:quantum_algo}, the cost of applying these unitaries scales polynomially with the system size. Overall, the quantum phase estimation algorithm applies the controlled $U$ operation a total of $\sum_{j=1}^t 2^j=2^{t+1}-2$ times to achieve precision $\varepsilon=2^{-n}=O(2^{-t})$. This means that the number of calls to a circuit implementing $U$ scales as $O(1/\varepsilon)$.

There are variants of quantum phase estimation that substitute the quantum Fourier transform by classical postprocessing and by iteratively refining the estimated phase~\cite{kitaev1995quantum}. Iterative methods have several advantages over standard quantum phase estimation, a notable one being that they are straightforward to parallelize. One of the best-performing examples is a Bayesian technique called rejection filtering phase estimation~\cite{wiebe2016efficient}. The empirical scaling of this method with respect to error is $4.7/\varepsilon$, close to the optimum of $\pi/\varepsilon$~\cite{berry2009perform}. 

In the next section, we describe the full quantum algorithm for estimating ground-state energies of cathode materials. This includes strategies for constructing the Hamiltonian, preparing approximate ground states as input, encoding Hamiltonians into unitaries, and performing quantum phase estimation.

\section{Quantum algorithm for battery simulation}
\label{sec:quantum_algo}

The quantum algorithm takes as input the Hamiltonian describing the interacting electrons in the material's unit cell and produces an estimate of its smallest eigenvalue, the ground-state energy. A method to represent and construct Hamiltonians is required, which should be tailored to the quantum algorithm. We define the electronic Hamiltonian in Sec.~\ref{ssec:hamiltonian} and explain why a first-quantization approach in a plane-wave basis~\cite{babbush2018low} is well suited for simulating battery materials. As described in Sec.~\ref{ssec:qalg}, the quantum phase estimation algorithm requires a method to prepare an approximate ground state to be used as input. This is challenging to perform both for periodic materials and in first quantization, so care must be taken to understand suitable methods for doing so. This is discussed in Sec.~\ref{ssec:initial_state}, where we outline the Hartree-Fock method for periodic materials and describe strategies for preparing the resulting Hartree-Fock state on a quantum computer. Finally, we outline how the qubitization formalism~\cite{low2019qubitization} can be used to encode the Hamiltonian into a suitable unitary. We employ the results of Ref.~\cite{su2021fault} to analyze the overall complexity of the algorithm and to compile all operations into a universal set of quantum gates compatible with fault-tolerant architectures. 

\subsection{First-quantized plane-wave Hamiltonians and wave functions}
\label{ssec:hamiltonian}

In the quantum phase estimation algorithm, there are three main choices to be made:

1. The Hamiltonian simulation technique used to encode Hamiltonians into unitaries. Widely-studied approaches include Trotterization~\cite{lloyd1996universal}, Taylor series~\cite{berry2015simulating}, qubitization~\cite{low2019qubitization} or interaction picture simulation~\cite{low2019hamiltonian}.

2. The state and Hamiltonian representation. This includes a choice between first or second quantization and potentially also a specific fermion-to-qubit mapping such as Jordan-Wigner~\cite{wigner1928paulische} or Bravyi-Kitaev~\cite{bravyi2002fermionic}.

3. The basis functions used to represent the state and the Hamiltonian. For material simulations, this is a choice between plane wave functions or localized atom-centered orbitals typically expanded in terms of contracted Gaussian functions~\cite{nagy2017basis, pritchard2019new}.

Plane waves are suited for the study of periodic systems and lead to compact representations of Hamiltonians. The challenge is that many plane waves are required to reach high accuracy, which leads to a prohibitively large number of qubits in second quantization. This motivates the choice of first-quantization techniques for materials simulation~\cite{su2021fault}. Finally, the qubitization approach, which is further described in Sec.~\ref{ssec:qubitization_qpe}, has the advantage that the desired unitary can be implemented exactly using a number of gates that scales linearly with the number of particles in the system, up to polylogarithmic factors~\cite{su2021fault}. Therefore, we focus on qubitization-based quantum phase estimation algorithms for first-quantized Hamiltonians represented in a plane-wave basis. We explain these concepts in more detail below.

The atomic structure of a cathode material is defined by its unit cell consisting of a group of atoms that can be translated in space to span the entire crystal. The electronic structure of the material can be obtained by solving the Schr\"odinger equation within the unit cell by imposing periodic boundary conditions. In the Born-Oppenheimer approximation~\cite{born1927quantum} the Hamiltonian describing the interacting electrons in the unit cell is given by
\begin{equation}
H = T + U + V,
\label{eq:h_total}
\end{equation}
where $T$ and $V$ are the kinetic energy and the electron-electron interaction operators defined by Eqs.~\eqref{eq:dft_T} and ~\eqref{eq:dft_V}, respectively, and $U$ is the Coulomb electron-nuclei interaction term given by
\begin{equation}
U = \sum_{i=1}^\eta \sum_{I=1}^L -\frac{Z_I}{\vert\vert \bm{r}_i - \bm{R}_I \vert\vert }.
\label{eq:h_en}
\end{equation}
In Eq.~\eqref{eq:h_en}, $L$ is the number of atoms in the unit cell, $Z_I$ is the atomic number of the $I$-th atom, and $\bm{r}_i$, $\bm{R}_I$ denote the positions of the electrons and nuclei, respectively.

Any complete set of basis functions can be used to represent the first-quantized Hamiltonian $H$. However, for periodic systems, using plane waves with the periodicity of the underlying lattice is a natural choice. More importantly, as we describe in Sec.~\ref{ssec:qubitization_qpe}, they significantly simplify the resulting expression for the Hamiltonian, which is beneficial to implement the quantum algorithm.

Plane-wave functions are defined as
\begin{equation}
\varphi_{p}(\bm{r}) = \frac{1}{\sqrt{\Omega}}e^{i \bm{G}_p \cdot \bm{r}},
\label{eq:pw_1}
\end{equation}
where $\Omega$ is the volume of the unit cell and the wave vector $\bm{G}_p$ is a reciprocal lattice vector (see App.~\ref{ssec:lattices}). For the case of an orthogonal lattice, we define
\begin{eqnarray}\label{eq:dfn_G_p}
&& \bm{G}_p = 2\pi \left[ \frac{p_1}{a_1}, \frac{p_2}{a_2}, \frac{p_3}{a_3} \right], \\
&& \bm{p} \in \mathcal{G}=\left[-\frac{N^{1/3}}{2}+1, \frac{N^{1/3}}{2}-1\right]^3,
\label{eq:pw_2}
\end{eqnarray}
where $a_1$, $a_2$, $a_3$ are the lattice constants, $N$ is the total number of plane waves, and the set $\mathcal{G}$ contains integer vectors defining a grid of points in the reciprocal lattice. Note that Eq.~\eqref{eq:pw_2} assumes a uniform distribution of points along the three orthogonal axes.

As described in more details in App.~\ref{app:hamiltonian}, by using the basis states $\ket{\bm{p}}$ such that $\braket{\bm{r}|\bm{p}}=\varphi_{p}(\bm{r})$, we can represent each term of the Hamiltonian as follows:
\begin{align}
T&=\sum_{i=1}^\eta\sum_{p\in \mathcal{G}}\frac{\|\bm{G}_p\|^2}{2}\ket{\bm{p}}\bra{\bm{p}}_i, \label{eq:t_op}\\
U&=-\frac{4\pi}{\Omega}\sum_{i=1}^\eta\sum_{q\in \mathcal{G}}\sum_{\substack{ \nu\in \mathcal{G}_0  \\ (\bm{q}-\bm{\nu}) \in \mathcal{G} }}\frac{\sum_{I=1}^L Z_I e^{i\bm{G}_\nu \cdot \bm{R}_I}}{\|\bm{G}_\nu\|^2}\ket{\bm{q-\nu}}\bra{\bm{q}}_i \label{eq:u_op},\\
V&=\frac{2\pi}{\Omega}\sum_{i\neq j}^\eta\sum_{p,q\in \mathcal{G}}\sum_{\substack{\nu\in \mathcal{G}_0 \\ (\bm{p} + \bm{\nu}) \in \mathcal{G} \\ (\bm{q}-\bm{\nu}) \in \mathcal{G}  } } \frac{1}{\|\bm{G}_\nu\|^2}\ket{\bm{p+\nu}}\bra{\bm{p}}_i \ket{\bm{q}-\bm{\nu}}\bra{\bm{q}}_j \label{eq:v_op}.
\end{align}
Here $\mathcal{G}_0$ is the set formed from $\mathcal{G}$ by removing the point $(0,0,0)$.

In first quantization, the wave function of $\eta$ electrons in a basis of $N$ single-particle wave functions (orbitals) is represented by directly specifying the single-particle state occupied by each electron: we employ $\eta$ registers each of size $n = \lceil \log N \rceil$, where the computational basis of each register enumerates the single-electron states. 

A general wave function for a system of interacting particles is written as a sum of weighted Slater determinants of $\eta$ electrons in $N$ orbitals:
\begin{equation}\label{eq:antisym-state}
    \ket{\psi} = \sum_{i \in {N\choose \eta}} c_i \mathcal{A}(\ket{\bm{p}_{i_1}, \ldots, \bm{p}_{i_\eta}}),
\end{equation}
where $\sum_i |c_i|^2=1$ with the index $i$ denoting a choice of $\eta$ occupied orbitals,  
\begin{equation}\label{eq:antisymmetrization}
\mathcal{A}: \ket{\bm{p}_{i_1}, \ldots, \bm{p}_{i_\eta}} \to \sum_{\sigma\in S_\eta}\frac{(-1)^{\pi(\sigma)}}{\sqrt{\eta!}}\ket{\sigma(\bm{p}_{i_1}, \ldots, \bm{p}_{i_\eta})},
\end{equation}
is the antisymmetrization operator, $S_\eta$ is the symmetric group on $\eta$ elements, $\pi(\sigma)$ is the parity of the permutation, and $\ket{\bm{p}_{i_1},\bm{p}_{i_2}, \ldots, \bm{p}_{i_\eta}}=\ket{\bm{p}_{i_1}}\ket{\bm{p}_{i_2}}\ldots\ket{\bm{p}_{i_\eta}}$ is an ordered product state of $\eta$ registers with $n$ qubits each. The number of qubits needed to represent the state scales logarithmically with $N$, requiring $3\eta \lceil \log(N^{1/3}+1)\rceil$ qubits, while a second-quantization approach would require $N$ qubits.

\subsection{Initial state preparation}
\label{ssec:initial_state}

As described in Sec.~\ref{ssec:qalg}, the quantum phase estimation algorithm requires an input state with sufficiently large overlap with the true ground state. In most quantum algorithms for quantum chemistry, this is done by preparing a state of non-interacting electrons described by single-particle wave functions (orbitals) that are optimized using the Hartree-Fock method~\cite{jensen2017introduction}. This state is usually referred to as the Hartree-Fock state. In second quantization, the Hartree-Fock state is straightforward to prepare since it is a computational basis state with no entanglement between qubits. The situation is more complicated when studying periodic materials in first quantization using a plane-wave basis. Here we need to apply the Hartree-Fock method to extended materials and provide an algorithm to prepare the resulting Hartree-Fock state in a plane-wave basis, which must be explicitly anti-symmetrized. We now describe how to perform each of these tasks.

\subsubsection{Hartree-Fock for periodic materials}\label{ssec:periodic-HF}
The Hartree-Fock method is a mean-field approximation that considers a state of independent electrons. These particles occupy orbitals that are optimized to minimize the energy of the state. In first quantization, they can be written as the anti-symmetric state $\mathcal{A}(\ket{\bm{p}_1, \bm{p}_2, \ldots, \bm{p}_\eta})$,
which is a special case of Eq.~\eqref{eq:antisym-state}. States of this form are referred to as Slater determinants.

The standard approach for obtaining the Hartree-Fock orbitals $\phi(\bm{r})$ is to express them as a linear combination of basis functions $\chi(\bm{r})$ as
\beq
\phi_i(\bm{r})=\sum_{\mu} C_{\mu i}\chi_\mu(\bm{r}),
\eeq
and optimize the coefficients $C_{\mu i}$. The optimal coefficients can be found by solving the generalized eigenvalue equations \cite{pople1954self, pople1992kohn, lehtola2020overview}
\beq\label{eq:RH}
FC=SCE,
\eeq
where $C$ is a coefficient matrix with entries $C_{\mu i}$, $F$ is known as the Fock matrix, $S$ is the overlap matrix, and $E$ is a matrix of eigenvalues. The Fock matrix and overlap matrix depend on integrals over the basis functions, as explained in Ref.~\cite{jensen2017introduction}. The most expensive step in the Hartree-Fock method is the construction of the Fock matrix and the overall complexity of the algorithm scales as $O(N^4)$~\cite{strout1995quantitative}. 

For periodic systems, it is important to ensure that the Hartree-Fock orbitals respect the periodicity of the system. From the Bloch theorem~\cite{ashcroft1976solid}, it follows that the wave function  describing the state of an electron in a periodic potential, e.g., the mean-field potential in a crystal structure, has the form
\begin{equation}
\phi_{n\bm{k}}(\bm{r}) = e^{i \bm{k} \cdot \bm{r}} u_{n\bm{k}}(\bm{r}),
\label{eq:bloch_state}
\end{equation}
where the wave vector $\bm{k}$ is the crystal momentum and the function $u_{n\bm{k}}(\bm{r})$ has the periodicity of the underlying crystal lattice. The allowed values of $\bm{k}$, known as k-points, are obtained by imposing the Born-Von Karman boundary condition on the wave function in Eq.~\eqref{eq:bloch_state}. See the App.~\ref{ssec:bands} for more details and for a summary of the general properties of electronic states in periodic potentials.

The function $u_{n\bm{k}}(\bm{r})$ can be expanded using the basis set of plane waves defined in Eq.~\eqref{eq:pw_1}, so the state $\phi_{n\bm{k}}(\bm{r})$ satisfies the Bloch theorem and is given by
\begin{equation}
\phi_{n\bm{k}}(\bm{r}) = \frac{1}{\sqrt{N_\mathrm{cell}\Omega}}\sum_\mu C_{\mu n}(\bm{k})  e^{i (\bm{k} + \bm{G}_\mu) \cdot {\bf r}},
\label{eq:pw_expansion}
\end{equation}
where $N_\mathrm{cell}$ is the number of unit cells in the macroscopic crystal. Eq.~\eqref{eq:RH} can be solved for each k-point to obtain the energies $E_n(\bm{k})$ and the optimized coefficients $C_{\mu n}(\bm{k})$ which define the electronic band structure of the material~\cite{liu2020simulating}. This is a feasible and straightforward approach, but the cost of the algorithm can become prohibitive if the number of plane waves $N$ is very large.

An alternative is to employ Bloch atomic orbitals as the basis functions~\cite{slaterKosterLCAO}. For an atom located at coordinates $\bm{R}_\mu$ with corresponding atomic orbitals $\chi_\mu(\bm{r})$, we define the Bloch atomic orbital
\beq
\chi_{\mu \bm{k}}(\bm{r})=\frac{1}{\sqrt{N_\mathrm{s}}}\sum_{i=1}^{N_s} e^{i\bm{T}_i\cdot \bm{k}}\chi_\mu(\bm{r}-\bm{T}_i-\bm{R}_\mu),
\eeq
where the sum runs over all $N_\mathrm{s}$ atomic sites and $\bm{T}_i$ is a lattice vector. The atom-centered  orbitals $\chi_\mu$ are typically approximated by a set of primitive Gaussian functions, which facilitates the calculation of the Fock matrix. The Hartree-Fock orbitals are then expressed as a linear combination of Bloch atomic orbitals
\beq\label{eq:HF-Bloch}
\phi_{n \bm{k}}(\bm{r}) = \sum_\mu C_{\mu n}(\bm{k}) \chi_{\mu \bm{k}}(\bm{r}).
\eeq
The advantage of this representation is that it is typically possible to work with much fewer Bloch atomic orbitals than plane waves to achieve a similar quality of the approximate ground state. Then, the Hartree-Fock equations can be solved using the localized basis set and transformed to the plane wave representation. It is important to note that quantifying the overlap between the Hartree-Fock state and the ground-state of periodic materials is an open problem of great importance since it directly affects the cost of quantum phase estimation.

\subsubsection{Antisymmetrization}\label{ssec:antisym}

First quantization and second quantization approaches differ in one important aspect: in the former case, the antisymmetry of fermionic systems appears in the state, while in the latter it does so in the operators. Therefore, it is important to describe how a quantum computer can antisymmetrize the initial state. This procedure will only need to be implemented once because the particle exchange operator commutes with the Hamiltonian, meaning that quantum phase estimation preserves the antisymmetry of the state.

The antisymmetrization procedure is mathematically defined in Eq.~\eqref{eq:antisymmetrization}.
Ref.~\cite{berry2018improved} introduced an efficient antisymmetrization algorithm that relies on the concept of a sorting network. The main idea is to employ the sorting network on an equal superposition state of auxiliary qubits, keep a record of the permutations made during the sort, and use the record to reverse the sorting operations on the state of the system. This results in an equal superposition of all permutations of the input state. The records can also be used to apply the corresponding phase to the permutation, resulting in the desired antisymmetrized state. More precisely, the algorithm of Ref.~\cite{berry2018improved} proceeds as follows (see Fig.~\ref{fig:Antisymmetrization}):

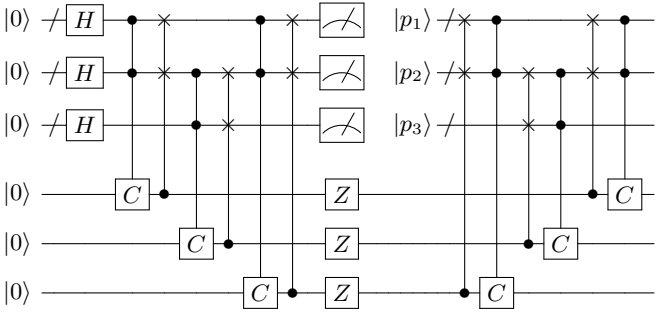
\begin{figure}
 \centering
 \Qcircuit @C=0.5em @R=0.75em { 
  & \ket{0} &  & & {/}\qw & \gate{H}  & \ctrl{1} & \qswap & \qw & \qw & \ctrl{1} & \qswap & \qw & \meter & & & &  \ket{p_1} & & & {/}\qw& \qswap & \ctrl{1}  & \qw & \qw & \qswap & \ctrl{1} & \qw \\
  & \ket{0} &  & & {/}\qw & \gate{H}  & \ctrl{3} & \qswap\qwx & \ctrl{1} & \qswap & \ctrl{5} & \qswap\qwx & \qw & \meter  & & & & \ket{p_2} & & & {/}\qw& \qswap\qwx & \ctrl{5}  & \qswap & \ctrl{1} & \qswap\qwx & \ctrl{3} & \qw \\
  & \ket{0} &  & & {/}\qw & \gate{H}  & \qw & \qw & \ctrl{3} & \qswap\qwx & \qw & \qw & \qw & \meter & & & & \ket{p_3} & & & {/}\qw& \qw & \qw  & \qswap\qwx & \ctrl{3} & \qw & \qw & \qw \\
  \\
  & \ket{0} &  & & \qw & \qw & \gate{C} & \ctrl{-3}  & \qw & \qw & \qw & \qw & \qw & \gate{Z} & \qw & \qw & \qw & \qw  & \qw & \qw  & \qw & \qw& \qw & \qw  & \qw  & \ctrl{-3} & \gate{C} & \qw \\
  & \ket{0} &  & & \qw & \qw  & \qw & \qw & \gate{C} & \ctrl{-4} & \qw & \qw  & \qw & \gate{Z} &  \qw & \qw & \qw  & \qw & \qw  & \qw & \qw & \qw & \qw & \ctrl{-3} & \gate{C} & \qw  & \qw & \qw \\
  & \ket{0} &  & & \qw & \qw& \qw & \qw & \qw & \qw & \gate{C} & \ctrl{-5} & \qw & \gate{Z} &  \qw & \qw & \qw  & \qw & \qw  & \qw & \qw  & \ctrl{-5} & \gate{C} & \qw & \qw & \qw  & \qw & \qw 
 }
\caption{\label{fig:Antisymmetrization}\textbf{Antisymmetrization circuit.} Example of an antisymmetrization circuit for three electrons. The operation $C$ represents a comparison test controlled on the two registers that are being compared. The seed register is measured in order to post-select on the collision-free subspace. The $Z$ gates perform the phase flip when swapping two registers. At the end of the circuit, the auxiliary record qubits (bottom register) can be discarded as they are disentangled~\cite{berry2018improved}. This circuit can be extended to an arbitrary number of electrons $\eta$ by increasing the size of the sorting network and adding additional auxiliary qubits for each required comparison and swap.}
\end{figure}

1. Define the function $f(\eta)=2^{\lceil\log(\eta^2)\rceil}\geq \eta^2$. Introduce an auxiliary seed system of $\eta$ registers each containing $\lceil\log(\eta^2)\rceil$ qubits. Apply a Hadamard gate on all qubits to create an equal superposition state
\beq
\frac{1}{\sqrt{f(\eta)^\eta}}\sum_{\ell_1, \ldots, \ell_\eta=0}^{f(\eta)}\ket{\ell_1, \ldots, \ell_\eta}.
\eeq

2. Introduce a record register containing as many qubits as there are sorting operations in the network and initialize it to the all-zero state. The state of the seed and record registers is
\beq
\frac{1}{\sqrt{f(\eta)^\eta}}\sum_{\ell_1, \ldots, \ell_\eta=0}^{f(\eta)}\ket{\ell_1, \ldots, \ell_\eta}\ket{0}.
\eeq

3. Apply the sorting network to the seed register and save the information on which swaps were made in the record register. The unnormalized state of seed and record after sorting is
\beq
\frac{1}{\sqrt{f(\eta)^\eta}}\sum_{0\leq \ell_1\leq\ldots\leq\ell_\eta}\left(\ket{\ell_1, \ldots, \ell_\eta}\sum_{\sigma\in S_\eta}\ket{\sigma_1, \ldots, \sigma_T}_\ell\right),
\eeq
where $\sigma_1 , \ldots , \sigma_T$ are the $T$ swaps applied by the sorting network that also decompose the permutation $\sigma = \sigma_1 \circ \cdots \circ \sigma_T$.

4. Project the state into the repetition-free subspace $\text{span}(\{\ket{\ell_1, \ldots, \ell_\eta} \,:\, \ell_i\neq\ell_j\}$). This projection is probabilistically done via a measurement of the seed register. The result is the state
\beq
\ket{\ell_1,\ldots, \ell_\eta}\frac{1}{\sqrt{|S_\eta|}}\sum_{\substack{\sigma\in S_\eta\\ \sigma = \sigma_1 \circ\cdots \circ \sigma_T }}\ket{\sigma_1, \ldots, \sigma_T}.
\eeq
If $f(\eta)$ is chosen as in step 1, the projection succeeds with probability greater than $1/2$~\cite{berry2018improved}. The seed register is disentangled from the system register and can be discarded.

5. Using the information in the record register, apply the inverse of the sorting network to the system with an additional $Z$ gate on each record qubit. Since the record register is in a superposition over permutations in $S_\eta$, this inverse sorting applies an equal superposition of all possible permutations of $\eta$ different elements, together with an overall phase corresponding to the parity of the permutation. We thus end up with an antisymmetric state, as desired.

The cost of this antisymmetrization procedure can be upper bounded by the use of up to three sorting networks: two for the sorting of the record register to account for the failure probability in the postselection, and the final one for inverse sorting on the system register. Calling $a = \lfloor \log \eta \rfloor$ and  $b = \lceil \log \eta\rceil$, the number $c$ of comparison operators in the network is $2^{a-1}a(a+1)/2\leq c\leq \lfloor n/2 \rfloor b(b+1)/2$. Each comparison operator can be implemented with $2\lceil (\log \eta) +1\rceil$ or $2\lceil \log N\rceil$ Toffoli gates~\cite{cuccaro2004new}, and half as many Toffolis are required to implement the controlled swap operations. Overall, the antisymmetrization can be performed using circuits with $O(\text{polylog}(\eta)\log\log N)$ depth.

\subsubsection{Preparing an arbitrary Slater determinant}\label{ssec:slater}
A common approximation to the ground-state is to consider an optimized Slater determinant defined at the $\Gamma$ point ($\bm{k}=(0,0,0)$) of the Brillouin zone (see App.~\ref{ssec:lattices}). Each independent electron occupies an orbital represented as a linear combination of basis functions, as captured by Eqs.~\eqref{eq:pw_expansion} and~\eqref{eq:HF-Bloch}. This means that in a plane-wave basis, the qubit representation of the first-quantized Hartree-Fock state, which is a single Slater determinant, is a superposition over the computational basis states.

Despite this state being more complicated than in second quantization, any single Slater determinant can be efficiently prepared on a quantum computer by performing transformations at the level of fermionic ladder operators~\cite{wecker2015solving,jiang2018quantum,kivlichan2018quantum}. This allows us to choose an initial basis of orbitals where the quantum states can be written as computational basis states and then perform a basis change into the optimized Hartree-Fock orbitals.

Any basis transformation of fermionic ladder operators can be described as
\begin{align}
    \tilde{a}^\dagger_{p}&= \sum_{q=1}^N u_{pq} a^\dagger_{q},\\
    \tilde{a}_{p}&= \sum_{q=1}^N u_{pq}^* a_{q},
\end{align}
where $a_{q}^\dagger,a_{q}$ are the ladder operators satisfying the canonical anticommutative relations, and $u_{pq}$ are entries of a $N\times N$ unitary matrix for a system with $N$ orbitals. Here $p,q$ are indices that indicate which orbitals are being considered. The entries of this unitary can be computed from the inner product between the initial $\varphi$ and final orbitals $\phi$
\begin{align}
    u_{pq} = \langle \phi_{q}, \varphi_{p}\rangle = \int d\bm{r} \phi_{q}^*(\bm{r})\varphi_{p}(\bm{r}),
\end{align}
where in our case $\varphi_{p}(\bm{r})$ corresponds to a plane wave defined by Eq.~\eqref{eq:pw_1} and $\phi_{q}(\bm{r})$ is a periodic Hartree-Fock orbital.

In the full Hilbert space, the transformation corresponds to a particle-preserving operation that can be written as
\beq
U(u) = \exp\left(\sum_{pq}\log[u_{pq}] (a_{p}^\dagger a_{q}-a_{q}^\dagger a_{p})\right).
\eeq
It was shown in Ref.~\cite{kivlichan2018quantum} that this transformation can be decomposed into a sequence of unitaries of the form
\beq
R_{pq}(\theta_{pq}) = \exp\left(\theta_{pq}(a_{p}^\dagger a_{q}-a_{q}^\dagger a_{p}))\right),
\eeq
and that these unitaries fulfill
\beq
R_{pq}(\theta_{pq}) U(u) = U(r_{pq}(\theta_{pq}) u),
\eeq
where $r_{pq}(\theta_{pq})$ is a Givens rotation on the two-dimensional subspace of row $p$ and column $q$~\cite[Sec.~11.3.1]{press2007vwt}, i.e., only entries $u_{pp}, u_{pq}, u_{qp}$ and $u_{qq}$ are affected by the rotation. It is possible to choose angles $\theta_{pq}$ such that the lower-triangle components of the matrix $u$ are set to zero by the rotation. Repeating this for each entry below the diagonal of the matrix effectively enacts a QR decomposition that diagonalizes $u$~\cite{kivlichan2018quantum}. This results in
\beq
\left(\prod_{p\neq q} R_{pq}(\theta_{pq})\right) U(u) = \prod_{p=1}^{N} e^{i \phi_{p} \hat{n}_{p}},
\eeq
which is a diagonal operator. Applying the inverse rotations directly gives a decomposition of $U(u)$:
\beq
 U(u) = \left(\prod_{p\neq q} R_{pq}(\theta_{pq})\right)^\dagger\prod_{p=1}^{N} e^{i \phi_{p} \hat{n}_{p}}.
 \label{Givens rotation operator}
\eeq

There are precisely $N \choose 2$ entries below the diagonal in matrix $U(u)$, but clever implementations can lead to a smaller number of rotations: since only rotations between $\eta$ occupied orbitals and $N-\eta$ unoccupied orbitals need to be performed, the actual number of rotations scales as $\eta(N-\eta)$. As we discuss in later sections, even though the cost of preparing this state is non-negligible, for practical simulations of cathode materials the cost of the full quantum algorithm is still dominated by quantum phase estimation. 

A further improvement is possible: in state preparation we can work with a number of plane waves $N'$ that is smaller than the one used for the full algorithm. The reason is that the error in approximating the ground state does not impact the final accuracy of the algorithm, but only its success probability by decreasing the overlap with the true ground state. Therefore, while the full algorithm is carried out on $n=\lceil\log N\rceil $, state preparation can be performed on a subset of $n'=\lceil\log N'\rceil$ qubits. This is important because it mitigates the linear scaling in $N$ of state preparation, ensuring that its final cost is smaller than the cost of running quantum phase estimation.

Now we show how to implement the Givens rotations. We begin by describing how this rotation acts in second quantization, derive its action in first quantization, and finally discuss its implementation algorithm. Consider a basis state in second quantization $a_{p_1}^\dagger \ldots a_{p_\eta}^\dagger \ket{\Omega}$, where the orbital indices $p_j$ are distinct and $\ket{\Omega}$ is the vacuum state. For this state, $R_{pq}(\theta_{pq})$ acts as the identity if none or both of $p,q$ are in $\{p_1,\ldots,p_\eta\}$. Otherwise, it applies the rotation 
\begin{equation}
R_{Y}(\theta_{pq}) = \begin{pmatrix}
\cos(\theta_{pq}) & \sin(\theta_{pq}) \\
-\sin(\theta_{pq}) & \cos(\theta_{pq})
\end{pmatrix},
\end{equation}
where the rows and columns correspond to $(a_{p}^\dagger,a_{q}^\dagger)$. For example, for $\eta=2$ and three orbitals $p,q,s$, we have $R_{pq}(\theta_{pq})a_{s}^\dagger a_{q}^\dagger\ket{\Omega} = \cos(\theta_{pq}) a_{s}^\dagger a_{q}^\dagger\ket{\Omega}  - \sin(\theta_{pq}) a_{s}^\dagger a_{p}^\dagger\ket{\Omega}$. 

A Givens rotation in first quantization must preserve the basis of antisymmetrized states $\mathcal{A}(\ket{p_1,\ldots,p_\eta})$. Consider the operator $\mathcal{B}: a_{p_1}^\dagger \ldots a_{p_\eta}^\dagger \ket{\Omega} \to \mathcal{A}(\ket{p_1,\ldots,p_\eta})$ that maps states in both representations. The transformation of interest in first quantization is then given by $\mathcal{B}R_{pq}(\theta_{pq})\mathcal{B}^\dagger$, which we denote by $R_{pq}^{(1)}(\theta_{pq})$. The previous example expressed in first quantization is then $R_{pq}^{(1)}(\theta_{pq})\mathcal{A}(\ket{s,q})= \cos(\theta_{pq}) \mathcal{A}(\ket{s,q})  - \sin(\theta_{pq}) \mathcal{A}(\ket{s,p})$.

To discuss the quantum circuit implementation, we extend the definition of $R_{pq}^{(1)}(\theta_{pq})$ to the full Hilbert space spanned by the states $\ket{p_1,\ldots ,p_\eta}$. The action on $\ket{p_1,\ldots, p_\eta}$ is the following. If $p_j\in \{p,q\}$ for exactly one $j$, it acts as the rotation $R_Y(\theta_{pq})$ on the subspace $\text{span}\{\ket{p},\ket{q}\}$ of the $j$-th register; otherwise it acts as the identity. This is indeed an extension of the operator by linearity. In our example, we have $R_{pq}^{(1)}(\theta_{pq}) \ket{s,q}= \cos(\theta_{pq})\ket{s,q}  - \sin(\theta_{pq}) \ket{s,p}$ and it is straightforward to compute $R_{pq}^{(1)}(\theta_{pq}) \ket{q,s}$, then subtract both states to derive the equation we wrote for $R_{pq}^{(1)}(\theta_{pq})\mathcal{A}(\ket{s,q})$.

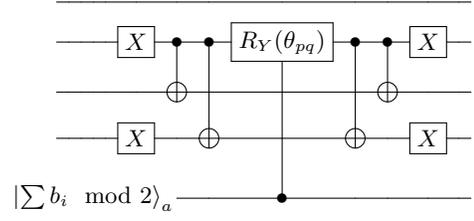
\begin{figure}
 \mbox{
 \Qcircuit @C=0.5em @R=0.85em { 
  &  & & & \qw& \qw& \qw & \qw & \qw & \qw & \qw & \qw & \qw & \qw & \qw &  \qw & \qw \\
  &  & & & \qw& \qw & \qw & \qw & \gate{X} & \ctrl{1} & \ctrl{2} & \gate{R_Y(\theta_{pq})} & \ctrl{2} & \ctrl{1} & \gate{X}& \qw& \qw \\ 
  &  & & &\qw& \qw & \qw & \qw & \qw & \targ & \qw & \qw & \qw & \targ & \qw & \qw & \qw  \\
  & & & & \qw& \qw& \qw & \qw & \gate{X} & \qw & \targ  & \qw & \targ & \qw & \gate{X}& \qw& \qw \\
  \\
    &   & & & & &\ket{\sum b_i \mod 2}_a   & &  &  & \qw  & \ctrl{-4} & \qw &  \qw & \qw &\qw &\qw  \\
 }}
 \caption{\label{fig:GivensControlled}\textbf{Circuit diagram of an example controlled rotation $R_Y(\theta_{pq})$}. The rotation is performed on the subspace spanned by $\ket{p} = \ket{0101}$ and  $\ket{q}=\ket{0010}$. The following procedure is applied to the bits where they differ, namely the last three qubits. First, we apply $X$ gates such that $\ket{p}\rightarrow \ket{0000}$ and $\ket{q}\rightarrow\ket{0111}$. Then, CNOT gates map these states to $\ket{0000}$ and $\ket{0100}$ respectively. This allows us to perform a rotation on the second qubit controlled on the auxiliary qubit $\ket{\sum b_i \mod 2}_a$. Finally, the CNOTs and $X$ gates are uncomputed, yielding the desired controlled rotation on the subspace $\text{span}\{\ket{p} , \ket{q}\}$.}
\end{figure}

We now describe how to implement the action of $R_{pq}^{(1)}(\theta_{pq})$ on $\ket{p_1,\ldots, p_\eta}$:
\begin{enumerate}
    \item Initialize $\eta$ auxiliary qubits in the state $\ket{0}_1 \ldots \ket{0}_{\eta}$. 
    \item For $1 \le j \le \eta$: If $p_j\in \{p,q\}$, flip the auxiliary qubit $\ket{0}_j$ to $\ket{1}_j$.
    \item For $1 \le j \le \eta-1$: Controlled on the auxiliary qubit $\ket{b_j}_j$, swap the $j$-th and $\eta$-th register.
    \item The auxiliary qubits are now in some state $\ket{b_1}_{1}\ldots\ket{b_{\eta}}_{\eta}$, where each $b_j$ indicates if $p_j\in \{p,q\}$. Controlled on the parity of $\sum_{i=1}^\eta b_i$, apply $R_Y(\theta_{pq})$ on the subspace $\text{span}\{\ket{p},\ket{q}\}$ of the $\eta$-th register. This step is illustrated for an example in Fig.~\ref{fig:GivensControlled} and can be easily generalized.
    \item Undo the controlled swaps and uncompute the auxiliary qubits by applying the same operators in steps 2 and 3.
\end{enumerate}
When $\ket{p_1,\ldots, p_\eta}$ contains none or both of $p,q$, we get $\sum b_i = 0 \pmod 2$. Thus, no rotation happens in step 4. After undoing the swaps and flips we return to the initial state, hence acting as the identity. Now assume $p_j \in \{p,q\}$ for exactly one $1\le j \le \eta$. Then $\sum b_i = 1 \pmod 2$, as $b_j=1$ and all others are zero. Thus the rotation is done on $p_j$ since $p_j$ will be located on the $\eta$-th register after step 3. After undoing the swaps, we get the desired state.

\subsection{Qubitization-based quantum phase estimation}
\label{ssec:qubitization_qpe}

We now focus on the most expensive part of the quantum algorithm, quantum phase estimation. 
The first step is to identify a method for encoding the Hamiltonian into a suitable unitary. Although there are several strategies to achieve this, the qubitization approach of Ref.~\cite{low2019qubitization} is particularly appealing because the resulting unitary can be implemented exactly without the need for any approximation. 

The qubitization-based encoding proceeds as follows.
A Hamiltonian can be written as a linear combination of unitaries 
\beq \label{eq:simple_LCU}
H = \sum_\ell \alpha_\ell H_\ell,
\eeq
where each $H_\ell$ is a unitary operator and we set $\alpha_\ell>0$, which can always be ensured by absorbing the phase inside the unitaries. The main strategy is to implement the operator $e^{-i \arccos(H)}$. This can be done exactly using the quantum walk operator \cite{berry2018improved}
\beq\label{eq:Q-operator}
Q=(2\ket{0}\bra{0}-I)\text{PREP}_H^{\dagger}\text{SEL}_H\text{PREP}_H,
\eeq
which acts on the system register and an additional auxiliary register. The prepare operator is defined as
\beq\label{eq:prep}
\text{PREP}_H\ket{0}\ket{\psi} = \left(\sum_\ell\sqrt{\frac{\alpha_\ell}{\lambda}}\ket{\ell}\right)\ket{\psi},
\eeq
where $\ket{\psi}$ is an arbitrary state and $\lambda=\sum_\ell \alpha_\ell$. The select operator is defined as
\beq\label{eq:sel}
\text{SEL}_H = \sum_\ell\ket{\ell}\bra{\ell}\otimes H_{\ell}.
\eeq

If $\ket{\Phi_k}$ is an eigenstate of $H$ with eigenvalue $E_k$, the operator $Q$ performs the transformation 
\beq
\begin{split}
Q\ket{0}\ket{\Phi_k} &=\frac{E_k}{\lambda}\ket{0}\ket{\Phi_k}-\sqrt{1-\left( \frac{E_k}{\lambda}\right)^2}\ket{\psi^\perp},
\end{split}
\eeq
where $\ket{\psi^\perp}$ is some state orthogonal to $\ket{0}\ket{\Phi_k}$. Defining $\cos(\theta_k)=\frac{E_k}{\lambda}$, a similar calculation can be performed to derive the action of $Q$ on $\ket{\psi^\perp}$, leading to the result:
\beq
\begin{split}
    Q\ket{0}\ket{\Phi_k}&=\cos(\theta_k)\ket{0}\ket{\Phi_k}-\sin(\theta_k)\ket{\psi^\perp},\\
    Q\ket{\psi^\perp}&=\cos(\theta_k)\ket{\psi^\perp}+\sin(\theta_k)\ket{0}\ket{\Phi_k}.
\end{split}
\eeq

The operator $Q$ is therefore block-diagonal, with each block $Q_k$ corresponding to a two-dimensional subspace $W_k=\text{span}\{\ket{0}\ket{\Phi_k}, \ket{\psi^\perp}\}$ that effectively forms a qubit, hence the name ``qubitization". Diagonalizing the two-dimensional submatrices $Q_{k}$ leads to a spectral decomposition~\cite{berry2018improved}
\begin{align} \label{eq:qwalk_eigenvalues}
Q_{k} &= e^{i\theta_k}\ket{\theta_k}\bra{\theta_k}+e^{-i\theta_k}\ket{-\theta_k}\bra{-\theta_k},
\end{align}
where $\ket{\pm \theta_k}$ are the eigenstates of $Q_k$. By using quantum phase estimation on $Q$ with an initial state close to the ground-state $\ket{0}\ket{\Phi_0}=\alpha \ket{\theta_k} + \beta \ket{-\theta_k}$ for some coefficients $\alpha$ and $\beta$, we obtain an estimate of $\theta_k$ with probability $|\alpha|^2$ and an estimate of $-\theta_k$ with probability $|\beta|^2$. Either result allows retrieval of the ground-state energy since $\cos(\pm \theta_k)=E_0/\lambda$.

In the standard formulation of quantum phase estimation, it is customary to apply the target unitary controlled on the state of an auxiliary qubit. However, it is also possible to instead apply the inverse unitary when the auxiliary qubit is in state $\ket{0}$~\cite{babbush2018encoding}. To do that, consider the reflection operator
\beq
R = \text{PREP}_H (2\ket{0}\bra{0}- I) \text{PREP}_H^\dagger.
\eeq
It satisfies the property
\beq\label{eq:W_conj}
R \;Q^n\; R= (Q^\dagger)^n.
\eeq
This implies that multiple controls on $Q^n$ can be replaced with a single control on whether $R$ operations are implemented before and after applying $Q^n$. From Eq.~\eqref{eq:Q-operator}, both $Q$ and $R$ can be implemented as sequences of select and prepare operators as well as the reflection operator $2\ket{0}\bra{0}-I$, for which standard circuit implementations are known~\cite{nielsen2002quantum}. Therefore, to determine how to implement the full quantum phase estimation algorithm, it suffices to specify how to implement the prepare and select operators. This is discussed in Sec.~\ref{ssec:Circuit_Implementation}.

To employ the qubitization approach for a fermionic Hamiltonian in first quantization, it is necessary to express it as a linear combination of unitaries. This was performed in Ref.~\cite{su2021fault} as follows. First assume that the lattice constants satisfy $a_1 = a_2 = a_3$ in Eq.~\eqref{eq:pw_2}, a restriction that we lift in App.~\ref{app:non_cubic}. The kinetic energy operator $T$ can then be simplified by observing that 
\begin{equation}
  \bm{G}_{p}^2= \left(\frac{2\pi \bm{p}}{\Omega^{1/3}}\right)^2 = \frac{4\pi^2}{\Omega^{2/3}}\sum_{\omega, r,s}2^{r+s}p_{\omega,r}p_{\omega,s},  
\end{equation}
where $p_{\omega,r}$ denotes the $r$-th bit of the $\omega$ component of the momentum vector. States with amplitudes defined by the product of momentum components are difficult to prepare, but they can be converted to phases by observing that $p_{\omega,r}p_{\omega,s} = \frac{1-(-1)^{p_{\omega,r}p_{\omega,s}}}{2}$. Thus, the operator $T$ in Eq.~\eqref{eq:t_op} can be rewritten as 

\begin{equation}\label{eq:T_operator}
\begin{split}
T&=\sum_{j=1}^\eta\sum_{w\in\{x,y,z\}}\sum_{r=0}^{n_p-2}\sum_{s=0}^{n_p-2}\frac{\pi^2}{\Omega^{2/3}}2^{r+s}\cdot\\
&\sum_{b\in\{0,1\}}\sum_{\bm{p}\in \mathcal{G}}\left((-1)^{b(p_{w,r}p_{w,s}\oplus 1)}\ket{\bm{p}}\bra{\bm{p}}_j\right),\\
\end{split}
\end{equation}
where $n_p=\lceil\log(N^{1/3}+1)\rceil$ is the number of qubits needed to store a signed binary representation of one component of the momentum vector. We can identify the amplitudes $\alpha_{\ell_T}$ and corresponding unitaries $H_{\ell_T}$ in the linear combination of unitaries expansion as
\begin{align}
  \alpha_{\ell_T} &= \frac{\pi^2}{\Omega^{2/3}}2^{r+s},\\\label{eq:H_ell_T}
  H_{\ell_T} &= \sum_{\bm{p} \in \mathcal{G}} (-1)^{b(p_{w,r}p_{w,s}\oplus 1)}\ket{\bm{p}}\bra{\bm{p}}_j,
\end{align}
where $\ell_T := (j, w, r, s, b)$.
Similarly, the operators $U$ and $V$ of Eqs.~\eqref{eq:u_op} and~\eqref{eq:v_op} can be rewritten as 
\begin{equation}\label{eq:U_operator}
\begin{split}
U&=\sum_{\bm{\nu}\in \mathcal{G}_0}\sum_{I=1}^{L}\frac{2\pi Z_I}{\Omega \|\bm{G}_{\nu}\|^2}\sum_{j=1}^\eta\sum_{b\in\{0,1\}}\\
&\sum_{\bm{q}\in \mathcal{G}}\left(-e^{i\bm{G}_{\nu}\cdot \bm{R}_I}(-1)^{b[(\bm{q}-\bm{\nu})\notin \mathcal{G}]}\ket{\bm{q}-\bm{\nu}}\bra{\bm{q}}_j\right),
\end{split}
\end{equation}
and
\begin{equation}\label{eq:V_operator}
\begin{split}
V&=\sum_{\bm{\nu}\in \mathcal{G}_0}\frac{\pi}{\Omega \|\bm{G}_{\nu}\|^2}\sum_{i\neq j=1}^\eta\sum_{b\in\{0,1\}}\sum_{\bm{p},\bm{q}\in G}\\
&\left((-1)^{b([\bm{p}+\bm{\nu}\notin \mathcal{G}]\vee[\bm{q}-\bm{\nu}\notin \mathcal{G}])}\ket{\bm{p}+\bm{\nu}}\bra{\bm{p}}_i\ket{\bm{q}-\bm{\nu}}\bra{\bm{q}}_j\right).
\end{split}
\end{equation}
Logical clauses such as $[\bm{p}+\bm{\nu}\notin \mathcal{G}]\vee[\bm{q}-\bm{\nu}\notin \mathcal{G}]$ indicate multiplication by one if the clause is satisfied and multiplication by zero otherwise. We identify the respective amplitudes and operators in the expansion as
\begin{align}
    \alpha_{\ell_U} &= \frac{2\pi Z_I}{\Omega \|\bm{G}_{\nu}\|^2},\\
  H_{\ell_U} &= \sum_{\bm{q} \in \mathcal{G}}-e^{i\bm{G}_{\nu}\cdot \bm{R}_I}(-1)^{b[(\bm{q}-\bm{\nu})\notin \mathcal{G}]}\ket{\bm{q}-\bm{\nu}}\bra{\bm{q}}_j,\\
  \alpha_{\ell_V} &= \frac{\pi}{\Omega \|\bm{G}_{\nu}\|^2},\\ \label{eq:H_ell_V}
  H_{\ell_V} &= \sum_{\bm{p},\bm{q} \in \mathcal{G}}(-1)^{b([\bm{p}+\bm{\nu}\notin \mathcal{G}]\vee[\bm{q}-\bm{\nu}\notin \mathcal{G}])}\ket{\bm{p}+\bm{\nu}}\bra{\bm{p}}_i\ket{\bm{q}-\bm{\nu}}\bra{\bm{q}}_j,
\end{align}
where we employ a similar indexing strategy $\ell_U:=(I,\bm{\nu},b,j)$ and $\ell_V:=(\bm{\nu},b,i,j)$. The phases $(-1)^{b[(\bm{q}-\bm{\nu})\notin \mathcal{G}]}$ and $(-1)^{b([\bm{p}+\bm{\nu}\notin \mathcal{G}]\vee[\bm{q}-\bm{\nu}\notin \mathcal{G}])}$ are designed to cancel out the amplitudes of the Hamiltonian terms where $\bm{p}+\bm{\nu}$ or $\bm{q}-\bm{\nu}$ lead to vectors outside of $\mathcal{G}$. For example, $b=0$ and $b=1$ give two opposite amplitudes for $\ket{\bm{q}-\bm{\nu}}\bra{\bm{q}}_j$ when $q-\bm{\nu} \notin \mathcal{G}$. Strictly speaking, the operators $H_{\ell_U},H_{\ell_V}$ are not unitaries, as their kernels include any state $\ket{\bm{q}-\bm{\nu}}$ or $\ket{\bm{p}+\bm{\nu}}$ outside of $\mathcal{G}$. Nevertheless, the qubitization procedure effectively implements these operators thanks to block-encoding, as explained later in App.~\ref{app:sel_operators}.

The parameter $\lambda=\sum_\ell \alpha_\ell$ can be viewed as a norm of the Hamiltonian. It is a crucial quantity in the complexity of quantum phase estimation because it effectively sets the energy scale of the problem: $E_0/\lambda$ has to be estimated with sufficient precision to recover $E_0$ with the desired error $\varepsilon$, which is challenging if $\lambda$ is large. Overall, this means $O(\lambda/\varepsilon)$ calls to a circuit implementing the qubitization operator are needed~\cite{berry2018improved}. The value of $\lambda$ can be calculated by adding the amplitudes for each component of the Hamiltonian~\cite{su2021fault}:
\begin{align}\label{eq:lambda_T}
\lambda_T &:=\sum_{\ell_T} \alpha_{\ell_T} =\frac{6\eta \pi^2}{\Omega^{2/3}}\left(2^{n_p-1}-1\right)^2 = O\left(\frac{\eta }{\Delta^2}\right),\\
\label{eq:lambda_U}
\lambda_U &:= \sum_{\ell_U} \alpha_{\ell_U}=\frac{\eta\sum_{I}Z_I}{\pi\Omega^{1/3}}\lambda_{\nu}= O\left(\frac{\eta^2}{\Delta}\right),\\
\label{eq:lambda_V}
\lambda_V &:= \sum_{\ell_V} \alpha_{\ell_V}=\frac{\eta(\eta-1)}{2\pi\Omega^{1/3}}\lambda_{\nu}= O\left(\frac{\eta^2 }{\Delta}\right),
\end{align}it is possible to gradually increase
where
\begin{equation}
 \lambda_\nu = \sum_{\bm{\nu}\in \mathcal{G}_0}\frac{1}{\|\bm{\nu}\|^2},\qquad
 \Delta = \left(\frac{\Omega}{N}\right)^{1/3}.
\end{equation}

A key result in Ref.~\cite{su2021fault} is to show that for fermionic Hamiltonians in first quantization, it is possible to implement the qubitization operator using circuits of depth $\Tilde{O}(\eta)$, where the tilde means that polylogarithmic terms are omitted. This leads to a total complexity $\Tilde{O}(\eta \lambda/\varepsilon)$ for the qubitization-based quantum phase estimation algorithm. More specifically, by setting $\Omega=O(\eta)$, meaning that the unit cell volume scales at most linearly with the number of particles,
the asymptotic complexity of the algorithm is
\beq
\Tilde{O}\left(\frac{\eta^{4/3}N^{2/3}+\eta^{8/3}N^{1/3}}{\varepsilon}\right).
\eeq
Moreover, taking constant resolution $\Omega=O(N)$, i.e., $\Delta=O(1)$, results in a scaling $\Tilde{O}(\eta^3/\varepsilon)$ that grows only polylogarithmically in $N$.

In the following, we outline in more detail how to implement the qubitization operator by focusing on the implementation of the prepare and select operators.

\subsection{Circuit implementation}\label{ssec:Circuit_Implementation}
As discussed above, quantum phase estimation is performed on the qubitization operator of Eq.~\eqref{eq:Q-operator}, which can be implemented in terms of the $\text{PREP}_H$ and $\text{SEL}_H$ operators of Eqs.~\eqref{eq:prep} and \eqref{eq:sel}. We follow the optimized compilation strategies pioneered in Ref.~\cite{su2021fault} to implement these operators. This section is therefore largely a concise summary of the results in Ref.~\cite{babbush2019quantum,su2021fault}.

To understand the strategy behind the implementation of those operators, we discuss qubitization at a more abstract level, and leverage the concept of block-encoding. For qubitization-based simulation to work, the operators $\text{PREP}_H$ and $\text{SEL}_H$ must satisfy certain properties~\cite{low2019qubitization}, chief among them the block-encoding identity
\begin{equation}\label{eq:block-encoding_id}
    \left(\bra{0} \text{PREP}_H^\dagger \cdot \text{SEL}_H \cdot\text{PREP}_H\ket{0}\right)\ket{\psi} = \frac{H}{\lambda}\ket{\psi}.
\end{equation}
Importantly, it is not required that these operators have the specific form in \eqref{eq:prep} and \eqref{eq:sel}; although those are preferred choices and can sometimes be realized. 

We then observe the following fact: if the linear sum of unitaries of $H$ decomposes to $H=A+B$, then the qubitization subroutines of $H$ can be defined in terms of those of $A$ and $B$. For example, given the decomposition $H = T + (U+V)$, we can define the state preparation subroutine of $H$ as 
\begin{align}\label{eq:prephdecomp}
    \resizebox{\hsize}{!}{$\left(\sqrt{\frac{\lambda_T}{\lambda}}\ket{0}+\sqrt{\frac{\lambda_U+\lambda_V}{\lambda}}\ket{1}\right)\otimes \PREP_T\ket{0}\otimes\PREP_{U+V}\ket{0}$},
\end{align}    
and the selection subroutine of $H$ as
\begin{align}\label{eq:selhdecomp}
    \ket{0}\bra{0}\otimes\SEL_T\otimes I+\ket{1}\bra{1}\otimes I\otimes\SEL_{U+V},
\end{align} 
and it is a straightforward calculation that the above definitions satisfy the block-encoding identity \eqref{eq:block-encoding_id}. Note the use of one additional qubit to distinguish the implementation of $T$ and $U+V$ subroutines. The same discussion applies to PREP$_{U+V}$ and SEL$_{U+V}$, which decompose into individual terms for $U$ and $V$. As a consequence, two qubits are needed to prepare a superposition with the amplitudes $\lambda_T$, $\lambda_U$, and $\lambda_V$.

\subsubsection{\label{sssec:Prep} Prepare operators}

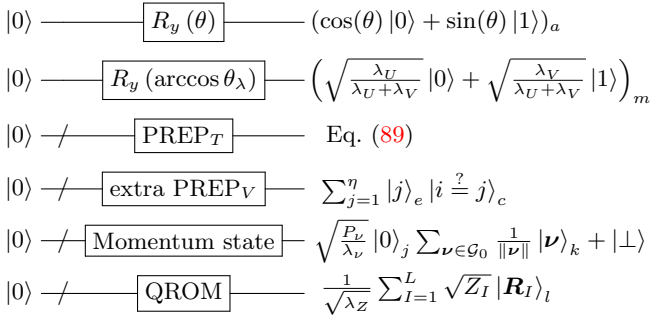
\begin{figure}
\Qcircuit @C=0.9em @R=0.8em {
 & \ket{0} & &\qw & \gate{R_y\left(\theta\right)} & \qw & & & & & & (\cos(\theta)\ket{0}+\sin(\theta)\ket{1})_a\\
  & \ket{0} & &\qw & \gate{R_y\left(\arccos \theta_\lambda\right)} & \qw & & & & & & & & \left(\sqrt{\frac{\lambda_U}{\lambda_U+\lambda_V}}\ket{0}+\sqrt{\frac{\lambda_V}{\lambda_U+\lambda_V}}\ket{1}\right)_m\\
  & \ket{0} & & {/}\qw 
  & \gate{\text{PREP}_T} & \qw & & & \text{Eq.~\eqref{eq:Prep_T_state_2}} \\
  & \ket{0} & & {/}\qw & \gate{\text{extra PREP$_V$}} & \qw& & & & & \sum_{j=1}^{\eta}\ket{j}_e\ket{i\overset{?}{=}j}_c\\
  & \ket{0} & & {/}\qw & \gate{\text{Momentum state}} & \qw & & & & & & & &  \sqrt{\frac{P_\nu}{\lambda_\nu}} \ket{0}_j\sum_{\bm{\nu}\in \mathcal{G}_0}\frac{1}{\|\bm{\nu}\|}\ket{\bm{\nu}}_k+\ket{\perp}\\
  & \ket{0} & & {/}\qw & \gate{\text{QROM}} & \qw & & & & & & \frac{1}{\sqrt{\lambda_Z}}\sum_{I=1}^{L}\sqrt{Z_I}\ket{\bm{R}_I}_l\\
}
\caption{\textbf{High-level representation of PREP subroutine}. The first two rotations can be attributed to the preparation of registers $a$ and $m$ in Eq.~\eqref{eq:prepsta}. The joint circuit for PREP$_T$ and the extra step of PREP$_V$ is shown in Fig.~\ref{fig:prep_t}. Finally, we have the preparation of the momentum state and the QROM routine.}
\label{fig:prep_h}
\end{figure}

We first explain how to prepare states with amplitudes $\sqrt{\frac{\alpha_{\ell_T}}{\lambda_T}}$, $\sqrt{\frac{\alpha_{\ell_U}}{\lambda_U}}$ and $\sqrt{\frac{\alpha_{\ell_V}}{\lambda_V}}$. This process is divided into four parts. A high level overview is given in Fig.~\ref{fig:prep_h}, along with a summary at the end of this section.\\

\begin{figure}
\mbox{
    \Qcircuit @C=0.9em @R=0.75em { 
  &\gate{H} & \ctrl{1} & \qw & \qw & \cdots&  & \qw & \gate{X}& \ctrlo{1} & \qw\\
  &\qw & \gate{H} & \ctrl{1}& \qw & \cdots &&\qw & \gate{X}& \ctrlo{1}&  \qw\\
 &\qw & \qw & \gate{H}&  \qw &\cdots & & \qw & \gate{X}& \ctrlo{0}& \qw\\
 \vdots & & & \vdots & &\ddots & & & \vdots & \vdots & \\
 & & & & & & & &\\
  &\qw & \qw & \qw&  \qw &\cdots & &\ctrl{1} & \gate{X}& \ctrlo{1}& \qw\\
 &\qw & \qw & \qw&  \qw &\cdots & &\gate{H} & \gate{X}& \ctrlo{1} & \qw\\
    & \ket{0}_{\text{flag}} &  &\qw & \qw & \qw & \qw & \qw & \qw &  \targ &  \qw
}    
}
\caption{\textbf{Preparation of a state proportional to  $\sum_r 2^{r/2}\ket{r}$, where $r$ is encoded in unary.} We prepare exponentially decreasing amplitudes by using a sequence of Hadamard and controlled-Hadamard gates. By flipping all qubits, the desired state with exponentially increasing amplitudes is obtained. The bottom qubit represents a flag qubit that can be measured to project out the invalid all-zero state.}
\label{fig:prep_r}
\end{figure}
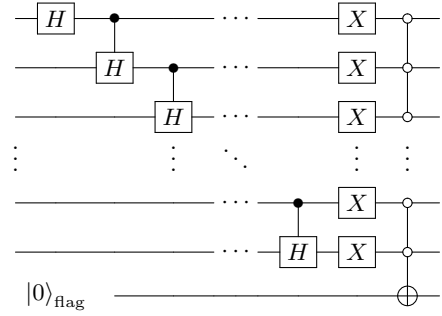

\begin{figure}[h]
\mbox{
\Qcircuit @C=0.9em @R=0.8em {
  & \ket{0} & & & &\qw     & \gate{H}               & \qw & \ket{+}      & & \qw & \qw \\
  & \ket{00}& & & & {/}\qw & \gate{\text{Uniform}}  & \qw & \ket{\omega} & & \qw & \qw\\
  & \ket{0}^{\otimes\log \eta} & & & & {/}\qw & \gate{\text{Uniform}}  & \qw & \ket{i} & & \ctrl{1} & \qw \\
  & \ket{0}^{\otimes \log \eta}& & & & {/}\qw & \gate{\text{Uniform}}  & \qw & \ket{j} & & \ctrl{1} & \qw \\
  & \ket{0}_{\text{flag}} & & & & \qw & \qw  & \qw & \qw & \qw & \gate{i = j} & \qw \\
& \ket{0}^{\otimes (n_p-1)} & & & & {/}\qw & \gate{\text{Exponential}}  & \qw & \ket{r} & & \qw & \qw \\
& \ket{0}^{\otimes (n_p-1)} & & & & {/}\qw & \gate{\text{Exponential}}  & \qw & \ket{s} & & \qw & \qw \\
}
}
\caption{\textbf{Circuit representation of PREP$_T$ and PREP$_V$}. The uniform superpositions can be implemented with a series of single control rotations or with Hadamard gates ($H$) and an inequality test. The creation of exponential superpositions of the form $\sum_r 2^{r/2}\ket{r}$ are implemented as in Fig.~\ref{fig:prep_r}. The preparation of the uniform superposition over $\ket{j}$ and the equality test with $\ket{i}$ is only required for PREP$_V$, but we include it here due to its conceptual similarity to other preparations in PREP$_T$.}
\label{fig:prep_t}
\end{figure}
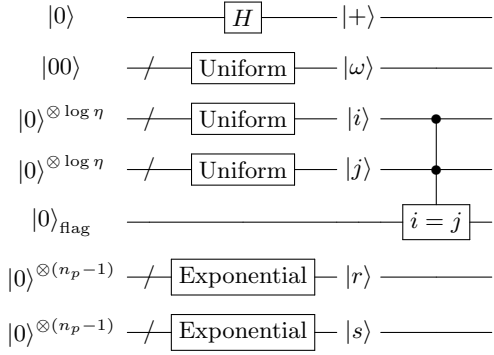

\paragraph{\label{par:PREP_T}Implementing PREP$_T$.}
We aim to prepare the state $\PREP_T\ket{0}=\sum_{\ell_T} \sqrt{\alpha_{\ell_T}/\lambda_T}\ket{\ell_T}$. By explicitly specifying all indices and assigning independent registers to each index, this state is proportional to
\begin{align} \label{eq:Prep_T_state}
2^{-(n_p-1)}\sum_{b,j,\omega, r,s}2^{(r+s)/2}\ket{b}_b\ket{j}_d\ket{\omega}\ket{r}_g\ket{s}_h,
\end{align}
or equivalently
\begin{equation}\label{eq:Prep_T_state_2}
\begin{split}
    2^{1-n_p}\ket{+}_b\sum_{j=1}^\eta\ket{j}_d\sum_{\omega=0}^2\ket{\omega}_f \sum_{r = 0}^{n_p-2}2^{r/2}\ket{r}_g \sum_{s = 0}^{n_p-2}2^{s/2}\ket{s}_h,
\end{split}
\end{equation}
where $\ket{+}=(\ket{0}+\ket{1})/\sqrt{2}$. A key step is to prepare a state of the form
\begin{equation}\label{eq:r_amplitudes}
    2^{-(n_p-1)/2}\sum_{r = 0}^{n_p-2}2^{r/2}\ket{r}.
\end{equation}
This can be performed with a circuit consisting of controlled Hadamard gates as shown in Fig.~\ref{fig:prep_r}. The overall circuit for implementing PREP$_T$ is illustrated in Fig.~\ref{fig:prep_t}.

\paragraph{\label{par:PREP_U+V}Momentum state for PREP$_{U+V}$.}
The challenge in preparing the target states  $\text{PREP}_{U}\ket{0}=\sum_{\ell_U} \sqrt{\frac{\alpha_{\ell_U}}{\lambda_{U}}}\ket{\ell_U},\text{PREP}_{V}\ket{0} = \sum_{\ell_V}\sqrt{\frac{\alpha_{\ell_V}}{\lambda_{V}}}\ket{\ell_V}$ is that the amplitudes $\alpha_{\ell_U},\alpha_{\ell_V}$ both depend on $\frac{1}{\|\bm{G}_{\nu}\|} = \frac{1}{2\pi}\frac{1}{\|\bm{\nu}\|}$. This common term means that the preparation of the momentum state below is required for both prepare operators:
\begin{equation} \label{eq:1/nu_state}
    \frac{1}{\sqrt{\lambda_{\nu}}}\sum_{\nu \in \mathcal{G}_0} \frac{1}{\|\bm{\nu}\|}\ket{\nu_x}\ket{\nu_y}\ket{\nu_z}.
\end{equation}

The detailed steps to prepare this state are summarized in App.~\ref{app:Momentum_state}, and we explain the high-level strategy here. We start by preparing a uniform superposition over acceptable values of $\ket{\bm{\nu}}$ and over auxiliary registers $\ket{m}$ and $\ket{\mu}$:
\begin{equation}\label{eq:1/nu_state_prep}
      \sum_{\mu=2}^{n_p+1}\sum_{\bm{\nu}\in B_{\mu}}\sum_{m=0}^{M} \frac{1}{2^\mu}\ket{\mu}\ket{\nu_x}\ket{\nu_y}\ket{\nu_z}\ket{m},
    \end{equation}
where $M$ should be judiciously chosen, and the sets $B_\mu := \{\nu \in \mathcal{G}_0 \,:\, 2^{\mu-2} \le \|\bm{\nu}\|_\infty < 2^{\mu -1}\}$ form a partition of $\mathcal{G}_0$. An inequality test is then used to discard part of the auxiliary state whenever $m \ge \lceil M(2^{\mu-2}/\|\bm{\nu}\|)^2\rceil$. As shown in App.~\ref{app:Momentum_state}, this corrects the amplitudes of $\ket{\bm{\nu}}$ as desired. 

The success probability of this procedure, dependent on the inequality test, converges to $0.2398$ with large $M$ and $n_p$~\cite[Eq.~(29)]{babbush2019quantum}. There are a few alternatives on how to deal with the failure case. One possibility is to first reduce its probability via amplitude amplification~\cite{brassard2002quantum} and then, in the rare case of failure, simply apply the identity operator. Since the effect of that is the addition of an identity to the Hamiltonian, we can correct any estimate made later on~\cite[Eq.~(30)]{babbush2019quantum}. Another solution is to use the failure case to apply the SEL$_T$ operator. This is done by modifying the subroutine equations and using a rotated auxiliary qubit $\cos(\theta)\ket{0}+\sin(\theta)\ket{1}$, where $\theta = \arcsin(2\sqrt{(\lambda_U+\lambda_V)/\lambda})$~\cite{su2021fault}. When $\lambda_T/(\lambda_U+\lambda_V) < 3$, the first approach decreases the leading term in the resource estimation formula (derived later in Eq.~\eqref{eq:cost_summarized}), resulting in a lower total cost in certain regimes, such as in our example application in Sec.~\ref{sec:application}.

Together with the steps described in the next parts, we use $\PREP'_{U+V}$ to denote the approximation to $\PREP_{U+V}$ that takes into account the failure probability.\\

\paragraph{\label{par:PREP_U_additional}Final step for PREP$_U$.}
To finish PREP$_U$, we prepare a new state $\frac{1}{\sqrt{\lambda_Z}}\sum_I \sqrt{Z_I}\ket{\bm{R}_I}$ where $\ket{\bm{R}_I}$ is a computational basis state encoding the position vector 
$\bm{R}_I$ and $\lambda_Z$ is a normalization factor. For this purpose, we employ a general state preparation technique called QROM~\cite{babbush2018encoding}, although other methods for preparing arbitrary states could be used since this leads to a small overhead. The state $\ket{\bm{R}_I}$ will be used later in $\text{SEL}_U$ to apply $-e^{i\bm{G}_{\nu}\cdot \bm{R}_I}$.\\

\paragraph{\label{par:PREP_V_additional}Final step for PREP$_V$.}
In contrast, PREP$_V$ requires creating an additional uniform superposition state proportional to $\sum_{i\neq j=1}^\eta \ket{i}_d\ket{j}_e$. To prepare such a state, we can use two uniform superpositions and a flag qubit to indicate whenever $i = j$.\\

\paragraph{Summary.}
 We collect all the states prepared so far. Depending on whether the value of $\lambda_T/(\lambda_U+\lambda_V)$ is greater or smaller than 3, the formula for the final state may change slightly due to some failure auxiliary flag. For the purpose of illustration, we assume $\lambda_T/(\lambda_U+\lambda_V) \ge 3$, which implies the following expression for the PREP subroutine application:
\begin{equation}
\left(\cos(\theta)\ket{0}+\sin(\theta)\ket{1}\right)
\otimes\PREP_T\ket{0}\otimes \PREP'_{U+V}\ket{0,0}.
\end{equation}

Consequently, the final state that we aim to prepare is~\cite[Eq.~(48)]{su2021fault}:
\begin{equation} \label{eq:prepsta}
\begin{split}
(\cos(\theta)\ket{0}+\sin(\theta)\ket{1})_a\ket{+}_b\left( \frac{1}{\sqrt{3}}\sum_{w=0}^2 \ket{w}_f\right)\\
\otimes\frac{1}{\sqrt{\eta}}\left( \sqrt{\eta-1} \ket{0}_c\sum_{i\neq j=1}^{\eta}\ket{i}_d\ket{j}_e+ 
\ket{1}_c\sum_{j=1}^{\eta}\ket{j}_d\ket{j}_e \right) \\
\otimes\left( \frac{1}{2^{n_p-1}-1}\sum_{r,s=0}^{n_p-2} 2^{(r+s)/2} \ket{r}_g \ket{s}_h \right)\\
\otimes\resizebox{\hsize}{!}{$\left(\sqrt{\frac{\lambda_U}{\lambda_U+\lambda_V}}\ket{0}+\sqrt{\frac{\lambda_V}{\lambda_U+\lambda_V}}\ket{1}\right)_m\left(\frac{1}{\sqrt{\lambda_Z}}\sum_{I=1}^{L}\sqrt{Z_I}\ket{\bm{R}_I}\right)_l$}\\
\otimes\left( \sqrt{\frac{P_\nu}{\lambda_\nu}} \ket{0}_j\sum_{\bm{\nu}\in \mathcal{G}_0}\frac{1}{\|\bm{\nu}\|}\ket{\bm{\nu}}_k+\sqrt{1-P_\nu}\ket{1}_j\ket{\bm{\nu}^\perp}_k \right),
\end{split}
\end{equation}
where $P_\nu$ is the probability of successfully preparing the momentum state. We have used different subscripts to denote different registers, which are explained below: 
\begin{enumerate}
\item $b,f,g,h$ are used for PREP$_T$.
\item $n,k$ are employed for the momentum state preparation, common to both PREP$_U$ and PREP$_V$.
\item $a$ is a rotated auxiliary register that allows us to apply SEL$_T$ when the momentum state preparation fails. 
\item $m$ is used for selecting between PREP$_U$ and PREP$_V$.
\item $l$ is exclusively used for PREP$_U$ in making the superposition $\frac{1}{\sqrt{\lambda_Z}}\sum_I \sqrt{Z_I}\ket{\bm{R}_I}$. 
\item $c,d,e$ are used for PREP$_V$, where $d,e$ each contain a superposition $\sum_{i=1}^\eta \ket{i}$, and $c$ is a flag qubit to indicate when $i = j$.
\end{enumerate}

The overall cost of implementing PREP$_H$ is dominated by the momentum state preparation, which includes the possible amplitude amplification procedure.

\subsubsection{\label{sssec:Sel} Select operators}

\begin{figure}
\mbox{
\Qcircuit @C=1.5em @R=1.25em { 
 & \ket{\bm{p}_1} && \qw &{/}\qw& \qswap  &\qw &\qw&\qw &\qw&\qw & \qswap&\qw\\
 & \ket{\bm{p}_2} && \qw &{/}\qw& \qw & \qswap &\qw &\qw& \qw & \qswap &\qw& \qw\\
 & \vdots & & & &\\
 &        & & & &\\
 & \ket{\bm{p}_{\eta}} && \qw &{/}\qw & \qw &\qw & \qswap & \qw &\qswap &\qw\qw &\qw& \qw\\
 &        & & & &\\
 &        & & & &\\
 &        & & & &\\
 & \ket{0} && \qw &{/}\qw& \qswap& \qswap& \qswap & \multigate{2}{O}& \qswap& \qswap& \qswap & \qw\\ 
  &        & & & &\\
 & \ket{a} && \qw &{/}\qw& \qw& \qw& \qw & \ghost{O}& \qw& \qw& \qw& \qw\\
 &        & & & &\\
 &        & & & &\\
 & \ket{j} && \qw &{/}\qw& \ctrl{-13} & \ctrl{-12} & \ctrl{-9} & \qw & \ctrl{-9} & \ctrl{-12} & \ctrl{-13}& \qw\\ 
  }
}
\caption{\textbf{Main technique used in the SEL operator.} The strategy used in SEL consists of (i) swapping the $\ket{\bm{p}_j}$ register into an auxiliary register, controlled on the value of $\ket{j}$; (ii) performing the target uncontrolled operation $O$, where some additional register $\ket{a}$ such as $\ket{\bm{\nu}}$ might intervene; and (iii) reversing the swaps. Each $\ket{\bm{p}_j}$ contains three $\omega$ coordinates $x$, $y$ and $z$, each with $n_p$ qubits. We use $O$ to represent different potential operations applied during SEL$_H$. In SEL$_T$, the operator $O$ represents the application of a phase $(-1)^{b(\bm{p}_{\omega,r}\bm{p}_{\omega,r} \oplus 1)}$. For SEL$_U$ and SEL$_V$, it may similarly refer to controlled phases or to arithmetic sums for computing $\ket{\bm{q-\nu}},\ket{\bm{p+\nu}}$.}
\label{fig:SEL_technique}
\end{figure}
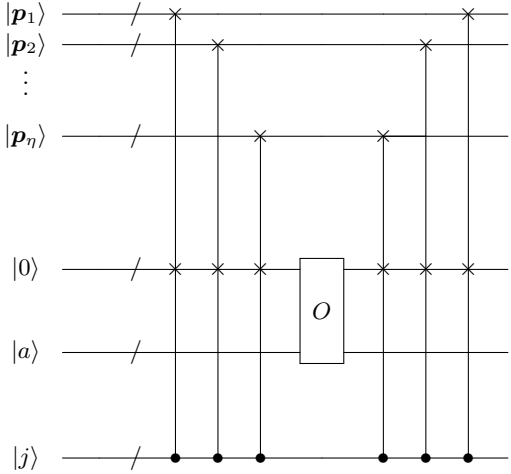

We now explain how to implement $\text{SEL}_T$, $\text{SEL}_{U}$ and $\text{SEL}_{V}$. The objective of these operators is to apply a unitary operation conditioned on the state of an auxiliary qubit. In our case, these unitaries either apply a phase or translate the corresponding momentum register by a given vector. The most straightforward way to apply SEL$_H$ is to iterate over the states $\ket{\bm{p}_j}$ and apply the corresponding unitary operators $H_\ell$ controlled on the state $\ket{j}$ of the auxiliary register. For example, an arbitrary operation $O$ can be applied as
\begin{align}
\begin{split}
    &\sum_{j=1}^{\eta}\ket{j}\bigotimes_{i=1}^{\eta} \ket{\bm{p}_i} \mapsto\\
    &\sum_{j=1}^{\eta}\Big(\ket{j} \otimes \ket{\bm{p}_1}\ldots O(\ket{\bm{p}_j})\ldots \ket{\bm{p}_{\eta}} \Big).
\end{split}
\end{align}
Each controlled application of $O$ on the right hand side acts on a different register. This is problematic because it requires many controlled operations. A more efficient technique consists of using control-SWAP operations to effectively transfer the target register to an auxiliary one and then performing uncontrolled operations on those registers before swapping them back~\cite[Eq.~(72)]{su2021fault}. We now explain this procedure in more detail. The circuit corresponding to this technique is shown in Fig.~\ref{fig:SEL_technique}. \\

\paragraph{\label{par:SEL_T}Implementing SEL$_T$.}
We implement a control-swap gate (CSWAP) of the $r$-th and $s$-th bits of the $\omega$-th component of $\ket{\bm{p}_j}$ into the state of auxiliary qubits, as follows. First, the state $\ket{\bm{p}_j}$ is swapped into an auxiliary register, controlled on the value of $\ket{j}_e$ (see Fig.~\ref{fig:SEL_technique}). 
\begin{align}
\begin{split}
    &\ket{j}_e\otimes \bigotimes_i\ket{\bm{p}_i}\otimes\ket{0}\mapsto\\
    &\ket{j}_e \otimes \Big(\ket{\bm{p}_1}\ldots\ket{0_j}\ldots\ket{\bm{p}_{\eta}}\Big)\otimes \ket{\bm{p}_j}.
\end{split}
\end{align}
Then we apply yet another CSWAP gate to the $\omega$-th component of $\ket{\bm{p}_j}$ to a new auxiliary register, controlled on $\ket{\omega}_f$, for $\omega\in\{x,y,z\}$. 
Finally, we perform similar CSWAPs moving the $r$-th and $s$-th bits of $\ket{\bm{p}_{j,\omega}}$ into two extra auxiliary qubits, controlled on $\ket{r}_g$ and $\ket{s}_h$, respectively. 

Eventually, this results in a copy of $\ket{\bm{p}_{j,\omega,r}}$ and $\ket{\bm{p}_{j,\omega,s}}$ in a register with two auxiliary qubits. Next, we use them as control of a controlled-controlled $Z$ gate with target register $b$, along with an additional $Z$ gate on $b$. The result is the phase $(-1)^{b\bm{p}_{j,\omega,r}\bm{p}_{j,\omega,s} + b}=(-1)^{b(\bm{p}_{j,\omega,r}\bm{p}_{j,\omega,s}\oplus 1)}$. This implements the target phase in Eq.~\eqref{eq:H_ell_T}. Finally, we use CSWAP gates to return the components of $\bm{p}$ into their original register.
Overall, the action of this circuit implementing the SEL$_T$ operator can be written as
\begin{equation}
\begin{split}
    &\text{SEL}_T:\ket{b}_b\ket{j}_e\ket{\omega}_f\ket{r}_g\ket{s}_h\ket{\bm{p}_j}\mapsto\\
    &(-1)^{b(\bm{p}_{j,\omega,r}\bm{p}_{j,\omega,s}\oplus 1)} \ket{b}_b\ket{j}_e\ket{\omega}_f\ket{r}_g\ket{s}_h\ket{\bm{p}_j},
\end{split}
\end{equation}
which makes use of the registers prepared in Eq.~\eqref{eq:Prep_T_state}.\\

\paragraph{\label{par:SEL_U}Implementing SEL$_U$ and SEL$_V$.}
The strategy for $U$ and $V$ is similar to the one described above for $T$ and illustrated in Fig.~\ref{fig:SEL_technique}. The main differences are that, from Eqs.~\eqref{eq:U_operator} and~\eqref{eq:V_operator}, to implement $\text{SEL}_{U}$ and $\text{SEL}_{V}$ we need to: (i) perform controlled sums and subtractions on the momentum registers, (ii) apply a phase to cancel out the amplitudes of the states where $\bm{p} + \bm{\nu} \notin \mathcal{G}$ or $\bm{q} - \bm{\nu} \notin \mathcal{G}$ (see the discussion after Eq.~\eqref{eq:H_ell_V}), and (iii) apply a second phase for the $-e^{i\bm{G}_{\nu}\cdot \bm{R}_I}$ exponent, which is only required for SEL$_U$. We describe these steps in further details in App.~\ref{app:sel_operators}.

The controlled swaps are the most expensive procedure in each step of quantum phase estimation; they are the only part of the implementation of the qubitization operator that scales linearly with the number of particles $\eta$. This is because we have to swap the $3\eta n_p$ qubits representing the system state (see Eq.~\eqref{eq:antisym-state}) in and out of the auxiliary qubits.

\section{Application: simulation of a cathode material}
\label{sec:application}
In this section, we estimate the resources required for running fault-tolerant quantum simulations of a cathode material using the quantum algorithm discussed in this manuscript. We start by providing a brief overview of the main families of cathode materials used in lithium-ion batteries. In Sec.~\ref{ssec:material} we focus on our use case: the polyanion material dilithium iron silicate ($\mathrm{Li}_2\mathrm{FeSiO}_4$)~\cite{eames2012insights}. Then, in Sec.~\ref{ssec:estimation} we report the qubit and gate costs as well as approximate runtimes for computing the ground-state energy of this material.

\subsection{Overview of cathode materials}
\label{ssec:cathode_materials}
In this section we provide a summary of the main characteristics of metal oxide cathodes which are discussed more extensively in Ref.~\cite{manthiram2020reflection}. Exploring new cathode materials has been crucial to improve the performance of batteries and to lower their cost. In particular, the groundbreaking discovery of metal oxide cathodes~\cite{goodenough2013li} allowed to significantly increase the operating cell voltage, and enabled the use of graphite anodes to overcome the safety problems associated with the use of lithium anodes \cite{cheng2017toward}. There are three classes of oxide cathodes that have been proposed for battery applications: layered oxides, spinel oxides, and the polyanion materials~\cite{manthiram2020reflection}. Among the different layered oxide materials with formula $\mathrm{LiMO}_2$, where M indicates a transition metal, the lithium cobalt oxide $\mathrm{LiCoO}_2$ has been a popular active material for commercial cathodes due to its high operating voltage, good structural stability and high ionic mobility \cite{mizushima1981lixcoo2, chevrier2010hybrid}. However, a large-scale deployment of the next-generation of lithium ion batteries will benefit from replacing cobalt with lower-cost and environmentally-friendly materials. 

This has motivated the exploration of novel mixed-metal layered materials with composition $\mathrm{LiNi}_{1-y-z}\mathrm{Mn}_y\mathrm{Co}_z\mathrm{O}_2$, the so-called NMC cathodes \cite{manthiram2020reflection}. These materials result from the progressive substitution of cobalt with the more abundant elements manganese and nickel. In these materials, manganese eases the incorporation of nickel while serving as a structure stabilizer. Furthermore, they exhibit a better chemical stability against oxygen loss from the cathode crystal lattice at highly delithiated phases \cite{jung2017oxygen}. In general, NMC cathodes show a high capacity across the full spectrum of compositions, which make them the leading cathode materials for automotive batteries~\cite{li2020high}. Spinel oxides have been also investigated \cite{thackeray1983lithium, spinel_cathodes_lu2016}. An advantage of the $\mathrm{LiMn}_2\mathrm{O}_4$ spinel oxide is the reduction in cost when compared with cobalt-based layered oxides. On the other hand, the number of chemical compositions for stable spinel-like phases is rather limited. Moreover, they are typically characterized by a lower cell voltage than traditional layered oxides~\cite{manthiram2020reflection}. 

The third class of cathode materials that have been investigated are the polyanion oxides~\cite{manthiram1989lithium, manthiram1989lithium}. Polyanion materials based on phosphates with composition $\mathrm{LiMPO}_4$ (M=Co, Ni) offer a promising avenue to increase the cell voltage to values as high as 5 volts~\cite{polyanion_materials}. On the other hand, the orthosilicates with stoichiometry $\mathrm{Li}_2\mathrm{MSiO}_4$ (M=Fe, Mn)~\cite{orthosilicate_cathodes_2008} have recently attracted significant attention. The landscape of possible materials opens also the possibility of using them to develop sodium-ion batteries~\cite{masquelier2013polyanionic}. Furthermore, an interesting feature of these materials is the possibility of extracting both lithium ions via a two-electron redox process, which could produce a higher capacity as compared to other cathodes~\cite{alternative_phosphate_2011}.

In the next section, we focus on the cathode material dilithium iron silicate $\mathrm{Li}_2\mathrm{FeSiO}_4$ oxide~\cite{orthosilicate_cathodes_2008}. This material is attractive in terms of sustainability since silicon and iron are among the most abundant elements on earth. Importantly, this silicate has a high thermal stability due to the strong covalent bond between the silicon and oxygen atoms~\cite{doughty2012general}. We have also selected this material as our use case because its conventional unit cell is orthogonal and significantly smaller than the unit cells of NMC cathodes. This facilitates the analysis and the implementation of the quantum algorithm.

\subsection{Analysis of the $\mathrm{Li}_2\mathrm{FeSiO}_4$ material}
\label{ssec:material}
Dilithium iron silicate belongs to the family of materials with tetrahedral structures~\cite{tetrahedral_structures} where the lithium, iron and silicon ions are coordinated by four oxygen atoms that form a tetrahedron. In general, these structures can be further classified into two families identified as $\beta$ and $\gamma$. In $\beta$-type structures, all tetrahedra point in the same direction. The $\gamma$ polymorphs instead self-assemble in groups of three, with the central tetrahedron oriented in the opposite direction to the outer two~\cite{islam2011silicate}.

\begin{figure}[h]
\includegraphics[width=1\columnwidth]{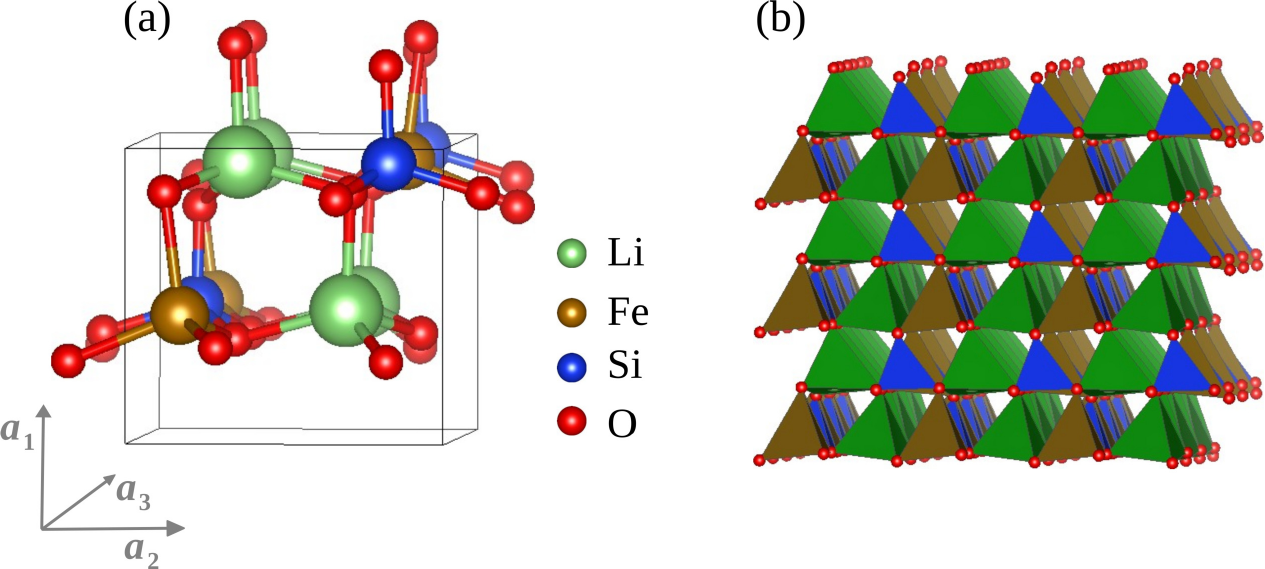}
\centering
\caption{(a) Conventional unit cell for the dilithium iron silicate $\mathrm{Li}_2\mathrm{FeSi}\mathrm{O}_4$ cathode material \cite{jain2013commentary}. (b) Crystal structure of the $\beta_{II}$ polymorph where all tetrahedra point in the same direction. Along the $\bm{a}_3$ direction, chains of $\mathrm{LiO}_4$ (green) are parallel to rows of alternating $\mathrm{SiO}_4$ (blue) and $\mathrm{FeO}_4$ (brown) tetrahedra representing the four-fold coordination of the lithium, silicon, and iron atoms by the oxygen atoms located at the vertices. This figure was produced using the VESTA package~\cite{vesta}.}
\label{fig:material}
\end{figure}

The conventional unit cell and the structure of the $\mathrm{Li}_2\mathrm{FeSiO}_4$ $\beta_{II}$-polymorph \cite{jain2013commentary} are shown in Fig.~\ref{fig:material}. The unit cell of this material is orthorhombic (stretched cubic lattice along two of its sides) and the crystal lattice is spanned by the primitive vectors $\bm{a}_1 = a_1(1, 0, 0), \bm{a}_2 = a_2(0, 1, 0), \bm{a}_3 = a_3(0, 0, 1)$, where $a_1=5.02$ $\mathrm{\AA}$, $a_2=5.40$ $\mathrm{\AA}$ and $a_3=6.26$ $\mathrm{\AA}$ are the lattice constants~\cite{eames2012insights}. The basis consists of sixteen atoms: four lithium (Li), two iron (Fe), two silicon (Si) and eight oxygen (O) atoms in the unit cell. The Li, Fe and Si ions are tetrahedrally coordinated by the oxygen atoms. From Fig.~\ref{fig:material}(b) we see that all tetrahedra point in the same direction, perpendicular to the close-packed planes. Moreover, along the $\bm{a}_3$ direction, the material consists of chains of $\mathrm{LiO}_4$ parallel to alternating rows of $\mathrm{SiO}_4$ and $\mathrm{FeO}_4$ tetrahedra.

The lithium intercalation into the silicate cathode material is represented by the chemical reaction

\begin{equation}
\mathrm{Li}_x\mathrm{FeSiO}_4 + (2-x)\mathrm{Li} \rightarrow \mathrm{Li}_2\mathrm{FeSiO}_4,
\label{eq:cathode_reaction}
\end{equation}
where $x$ in the equation above indicates the number of lithium atoms that have been removed per formula unit. The analogue of Eq.~\eqref{eq:voltage_DE} for this cathode material can be used to compute the equilibrium cell voltage
\begin{equation}
V = -\frac{ \left[ E_{\mathrm{Li}_2\mathrm{FeSiO}_4} - E_{\mathrm{Li}\mathrm{FeSiO}_4} - E_\mathrm{Li} \right]}{F},
\label{eq:voltage}
\end{equation}
which involves the energy difference between the lithiated ($\mathrm{Li_2FeSiO}_4$) and delithiated ($\mathrm{LiFeSiO}_4$) phases of the material, where the latter is produced by removing one lithium atom per formula unit.

In practice, the total energies entering Eq.~\eqref{eq:voltage} are typically obtained from first-principles density functional theory (DFT) calculations. Previous DFT simulations for this material have underestimated the experimental voltage ($\sim 3.10$ volts) by roughly $0.4-0.7$ volts~\cite{nyten2006lithium, larsson2006ab, kokalj2007beyond}. The large deviations have been ascribed to the self-interaction error of semi-local functionals and the lack of error cancellations in Eq.~\eqref{eq:voltage}, as discussed in Sec. \ref{sec:dft}. Authors in Ref.~\cite{eames2012insights} have used the DFT+U correction to predict a more accurate voltage which is $0.24$ volts above the experimental value~\cite{nyten2006lithium}.

\subsection{Resource estimation}
\label{ssec:estimation}

In this section, we discuss the gate cost, qubit cost, and estimated runtime of implementing the quantum algorithm for calculating the ground-state energy of the $\mathrm{Li_2FeSiO}_4$ cathode material. All calculations to derive these costs have been carried out using the \href{https://github.com/PabloAMC/TFermion/tree/first_quantization}{TFermion} library~\cite{casares2021t}.

\subsubsection{Gate cost}\label{sssec:Gate_cost}
In the setting of fault-tolerant quantum computing, it is customary to distinguish between Clifford gates, which satisfy symmetry properties that make them easier to implement, and non-Clifford gates, which are much more expensive and therefore carry the leading cost of the quantum computation~\cite{fowler2013bridge}. Typically T gates or Toffoli gates are the non-Clifford gates considered in practical error-correcting codes such as the surface code~\cite{fowler2012surface}. Non-Clifford gates are expensive because they cannot be transversely and fault-tolerantly implemented in two-dimensional codes \cite{kitaev2003fault,bombin2006topological} --- the codes with the most favorable thresholds. Consequently, their fault-tolerant implementation requires either `code switching' \cite{bombin2016dimensional} to three-dimensional codes~\cite{bombin2007topological,vasmer2019three} for the T or Toffoli gates, or a process known as magic state distillation which usually has a lower overhead~\cite{beverland2021cost}. Magic state distillation can however require many physical qubits and rounds of error detection within such codes, during which Clifford operations can be applied in parallel~\cite{babbush2019quantum}. 

The most expensive step of the algorithm is performing qubitization-based quantum phase estimation. The central result of Ref.~\cite{su2021fault} is an explicit and general formula for the number of Toffoli gates required to implement this algorithm for a first-quantized Hamiltonian. For clarity, we have reproduced it fully in Eq.~\eqref{eq:cost} in the appendix, and summarize it here by writing the leading terms. By taking only these leading terms, the number of Toffoli gates required is equal to
\begin{equation}\label{eq:cost_summarized}
    \left\lceil \frac{\pi \lambda}{2\varepsilon_{\text{QPE}}}\right\rceil \left(12\eta n_p + \text{polylog}(\eta, N, \varepsilon) + \lambda_Z + Er(\lambda_Z) \right),
\end{equation}
where we recall $n_p = \lceil \log (N^{1/3}+1)\rceil$ is the number of qubits needed to represent a component of the signed plane-wave index, $\lambda_Z= \sum_I Z_I$ is the sum of nuclear charges, $\varepsilon_{\text{QPE}}$ is the accuracy of quantum phase estimation, and $\text{Er}(x)=\min_m (2^m+\lceil 2^{-m}x\rceil)$. The dominant term in the cost of the algorithm is the prefactor $\lambda$, which depends on the number of particles $\eta$, the success probability of the momentum state preparation, and on the spacing parameter $\Delta = \left( \frac{\Omega}{N}\right)^{1/3}$ in the cell. This $\lambda$ is not exactly $\lambda_T+\lambda_U+\lambda_V$, but has to be slightly increased to take into account some failure probabilities and implementation decisions; for a more detailed discussion see Ref.~\cite[Eqs.~116-124]{su2021fault}. In the case of $\mathrm{Li_2FeSiO}_4$, the cell consists of $\eta = 156$ electrons and has dimensions $5.02\times 5.40\times  6.26 \mathrm{\AA}^3$, amounting to $\Omega \approx 1145 a_0^3$, where $a_0$ is the Bohr radius. 

The number of plane waves $N$ is a free parameter of the algorithm, which can be chosen to achieve a desired basis error $\varepsilon_b$ in representing wave functions. This error scales as $\varepsilon_b = \Tilde{O}(1/N)$~\cite[App.~E]{babbush2018low}. While it is difficult to provide the prefactors required to quantify the basis error exactly as a function of $N$, a variety of heuristic guidelines can be employed. It has been argued that in periodic materials roughly 10-20 times as many plane waves as Gaussians are needed for the same level of precision~\cite{babbush2018low}. Taking the Dunning basis sets from cc-pVDZ to cc-pV5Z as a comparison point \cite{dunning1989gaussian}, we get $10^4$ to $10^6$ plane waves. We can also take the inverse density of plane waves as a point of reference:  $\Delta = 10^{-2} a_0$ is expected to be more accurate than large Gaussian basis sets~\cite{su2021fault}. Taking ranges from $\Delta \in [10^{-2}a_0, a_0]$ translates to $N \in [10^3,10^9]$. Another free parameter of the algorithm is the target precision $\varepsilon_{\text{QPE}}$ for quantum phase estimation. The error in the phase estimation can be directly linked to an error in the voltage estimation using Eq.~\eqref{eq:voltage}.  

\begin{figure*}
\includegraphics[width=.47\textwidth]{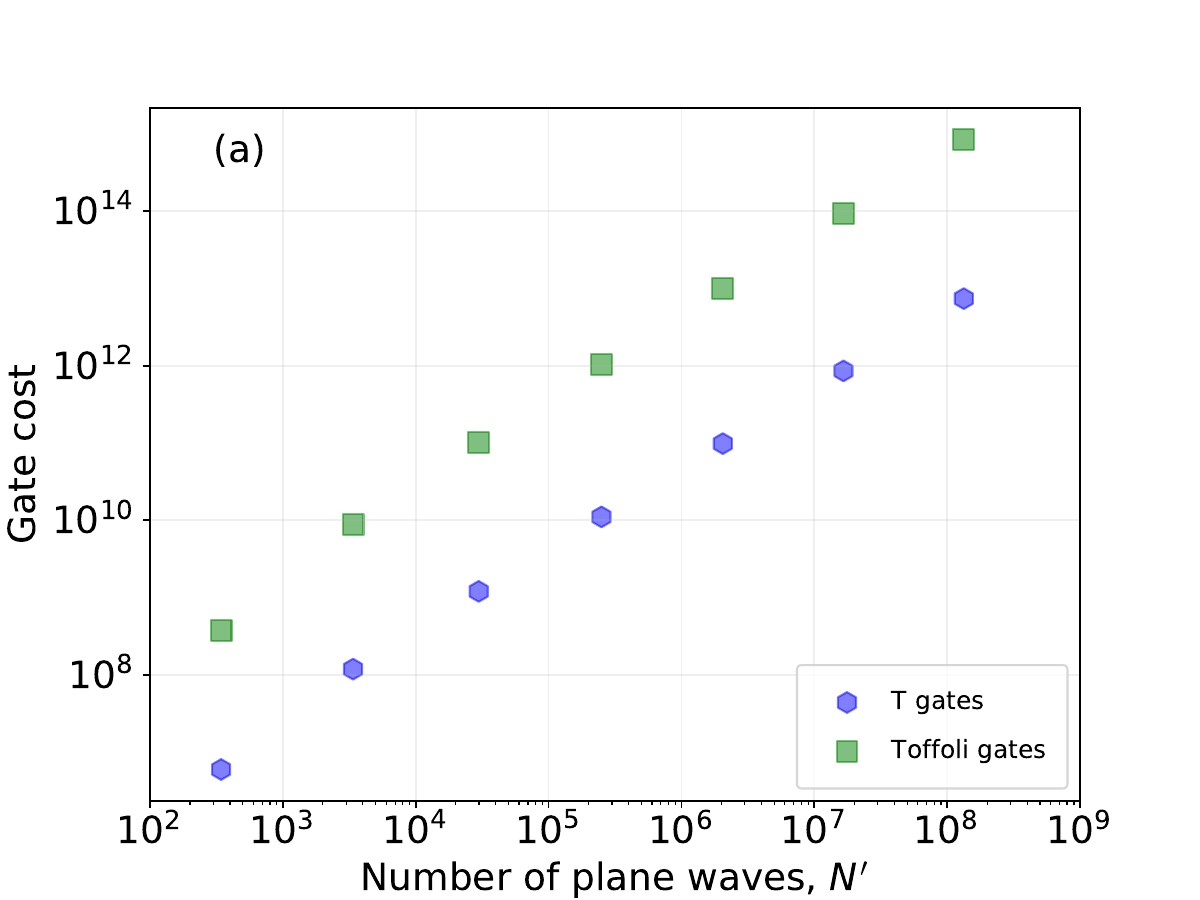}
\includegraphics[width=.47\textwidth]{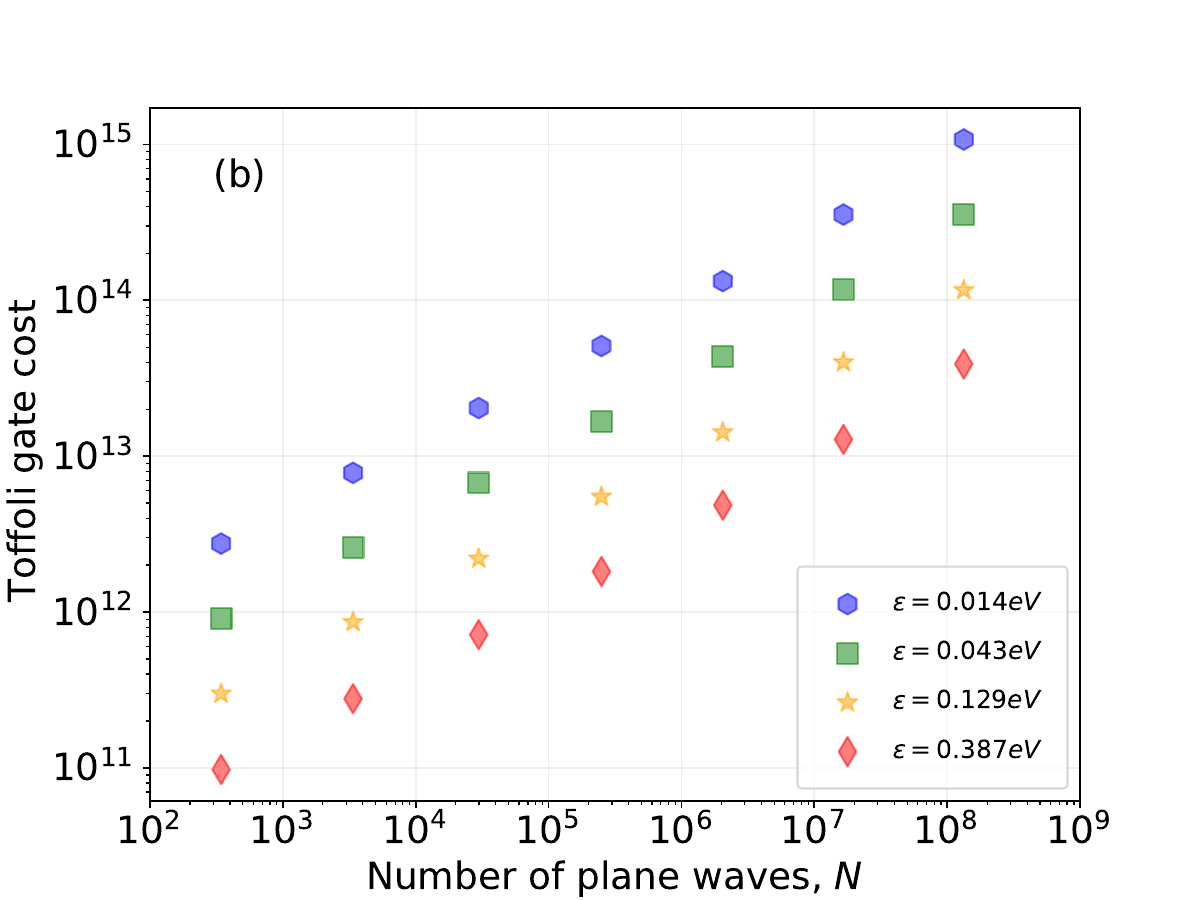}
\caption{\label{fig:cost}\textbf{Non-Clifford gate cost for initial state preparation and quantum phase estimation}. (a) The non-Clifford gate cost due to Givens rotations used in the circuit for initial state preparation. (b) Toffoli gate cost of the quantum phase estimation algorithm. All calculations are done for the unit cell of Li$_2$FeSiO$_4$ with 156 electrons. The total number of qubits is 2,375 for $n_p=4$ and 6,652 for $n_p=9$. In the right figure we only depict Toffoli gate count, as the number of T gates is much smaller ($<3\times 10^{5}$). The total error $\varepsilon$ includes contributions from different approximations throughout the algorithm, but it does not take into account the error derived from a finite basis set. The slope of the Toffoli gate cost for fixed target precision is a consequence of the leading cost term in \eqref{eq:cost_summarized}, $12 \eta n_p\left\lceil (\pi\lambda)/(2\varepsilon_{QPE}) \right\rceil$, where $n_p = \lceil \log (N^{1/3}+1)\rceil$. These calculations were performed with the T-Fermion library~\cite{casares2021t}.}
\end{figure*}

Fig.~\ref{fig:cost} represents how the cost of the full algorithm, captured by the number of Toffoli gates, depends on different values of the number of plane waves $N$ and the error in the ground-state energy estimation. The discrete number of qubits required to represent the quantum state, and its direct relation to the Toffoli cost \eqref{eq:cost_summarized}, suggest using a number of plane waves $N$ translating directly to integer values of $n_p$. Therefore in our resource estimations, we take $N = (2^{n_p}-1)^3$ for $n_p\in[3,9]$. These resource estimations differ in two details from the full gate cost equation presented in Eq.~\eqref{eq:cost}. First, the phase estimation error $\varepsilon_{\text{QPE}}$ is not the only error source we consider; we also take into account others due to the finite number of bits used to represent $\ket{m}$ or $\ket{\bm{R}_I}$ in the PREP$_U$ implementation. We refer to these as $\varepsilon_M$ and $\varepsilon_R$ respectively. Additionally, we include the modification needed for dealing with a non-cubic unit cell, as explained in App.~\ref{app:non_cubic}.

There is also one subtlety that explains why we only count Toffoli gates in the phase estimation algorithm. In principle, there is an important contribution of T gates from the rotations required to implement $e^{-i \bm{G}_{\nu}\cdot \bm{R}_I}$, which should be implemented over all applications of the qubitization operator. However, there is a way to avoid such T gates, as shown in Ref.~\cite{su2021fault}. The key idea is to perform the addition in the dot product $\bm{G}_{\nu}\cdot \bm{R}_I$ on a gradient phase state $2^{-b/2}\sum_{k=0}^{2^b-1} e^{-2\pi ik/2^b}\ket{k}$. This procedure, via a phase kickback \cite{gidney2018halving,kitaev2002classical}, implements the desired rotations without increasing the number of T gates that are required to prepare the gradient phase state in the first place. Finally, additional T gates might be needed in the inverse QFT, but those are negligible.

\subsubsection{Qubit cost}\label{sssec:Qubit_cost}
The full description of each contribution to the total number of logical qubits for the quantum phase estimation algorithm is given in~\cite[App.~C]{su2021fault}. This can be applied to the cathode material we study as the modifications required to accommodate a non-cubic lattice~(App.~\ref{app:non_cubic}) change the qubit cost by a relatively small constant. We do not reproduce here these logical qubit numbers, but instead give their overall sum:
\begin{equation}
\begin{split}
    3\eta n_p + 4n_Mn_p + 12n_p + 2\left\lceil \log\left(\left\lceil\frac{\pi\lambda}{2\varepsilon_{\text{QPE}}}\right\rceil\right)\right\rceil  +\\
    2\lceil \log(\eta)\rceil+ 5n_M + 3n_p^2+ \lceil \log(\eta+2\lambda_Z)\rceil +\\
    \max(5n_p+1,5n_R-4)+\max(n_T, n_R+1)+ 33,
\end{split}
\end{equation}
where $n_M,n_R,n_T$ represent numbers of qubits that are determined according to non-Clifford gate optimization of the different error sources of the algorithm. These include for example the choice of $M$ in the momentum state preparation as in Eq.~\eqref{eq:1/nu_state_prep} and the number of bits to represent nuclei coordinates. For each quantity $n_M,n_R,n_T$, the number of qubits ranges from $30$ to $50$ depending on the total error budget, while $\lceil \log(\eta+2\lambda_Z)\rceil\approx 9$. Note however that the leading term is $3 \eta n_p$, corresponding to the $\eta$ momenta registers, each using $n_p$ qubits for three coordinates. 

In contrast, during the initial state preparation, we need $3\eta n_p$ qubits to represent the quantum state, and $(3n_p-1) +1$ auxiliary qubits for the multi-control NOT operations.
With these choices, the overall number of logical qubits is 2,375 for $n_p=4$ and 6,652 for $n_p=9$, where the leading term $3 \eta n_p$ provides the most significant contribution of 1,872 and 4,680 logical qubits, respectively.

\subsubsection{\label{sec:runtime}Algorithmic runtime}
While we leave an accurate runtime analysis for future work, rough estimates can be obtained. There are three main variables that determine the runtime: the surface code distance $d$ (correcting $\lfloor d/2\rfloor$ errors in computations), the number of non-Clifford gates $N_{nc}$, and the clock rate $f$ for applying gates. The total time is then given by $N_{nc}d/f$.

First, we need to obtain the surface code distance $d$ leading to a logical failure rate low enough to perform the computation. These calculations depend on the hardware platform, and for this purpose we focus on photonic architectures such as those described in Refs.~\cite{tzitrin2021fault,kim2021fault}. More specifically, we employ the formula in~\cite[Eq.~9]{kim2021fault} for the value of $d$ in our estimations, which leads to values of $d$ between 30 to 40.

Next, we discuss the number of non-Clifford gates. As mentioned before, non-Clifford gates are produced using distillation of magic states made in so-called magic state factories. In algorithms with few qubits, these factories make up a large percentage of the quantum computer. However, for an algorithm with qubit costs in the thousands, the footprint is rather small (roughly $2\%$). Thus, parallelization techniques~\cite{low2018trading,campbell2017unified} that do not depend on the hardware can enable fast injection of distilled magic states, reducing runtime by an order of magnitude. In a further optimization, since the algorithm we study relies heavily on Toffoli gates (Fig.~\ref{fig:cost}), we can directly synthesize them using efficient magic state factories~\cite{gidney2019efficient}. This generally quintuples the speed of the algorithm compared to previous state-of-the-art procedures~\cite{fowler2019low}.

\begin{figure}
\includegraphics[width=1\columnwidth]{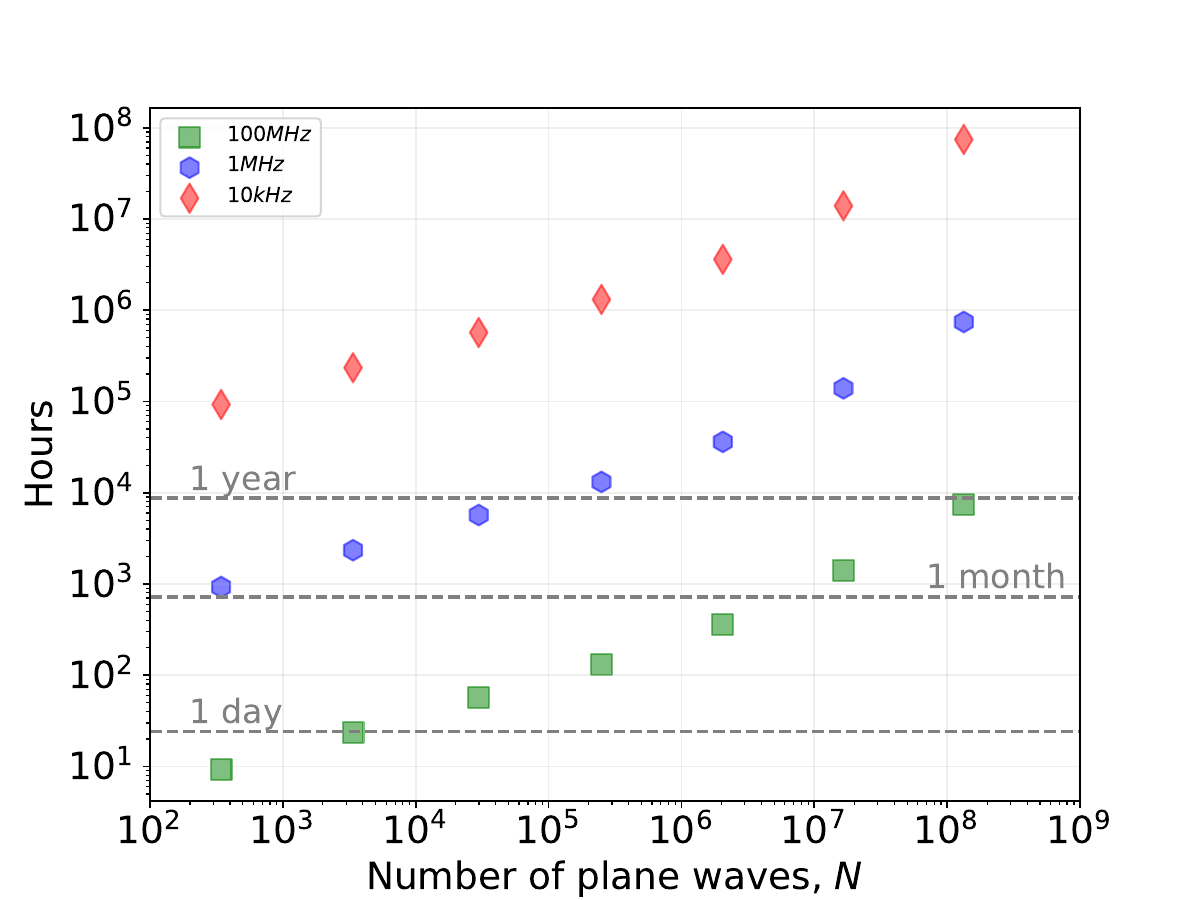}
\centering
\caption{\textbf{Estimation of the time required to run the algorithm}. This figure illustrates total runtime for synthesizing all the Toffoli gates indicated in Fig.~\ref{fig:cost} for $\varepsilon = 0.043$ eV.  All calculations are done for the unit cell of Li$_2$FeSiO$_4$ with 156 electrons, and we assume that the number of plane waves used in the state preparation and quantum phase estimation are the same. The total number of qubits is 2,375 for $n_p=4$ and 6,652 for $n_p=9$. We compute the distillation time as the product of the number of Toffoli gates, the surface code distance $d$, and the clock frequency, all divided by a small $n_p$ factor originating from the techniques in~\cite{low2018trading} that parallelize the CSWAPs and arithmetic computations. We compute $d$ in this figure as in the moderate error case of Ref.~\cite{kim2021fault}. We emphasize that these are rough estimates whose main purpose is to provide a method to interpret the gate cost.}
\label{fig:time}
\end{figure}

Finally, we discuss the clock rate, which is the most challenging value to calculate, with there being a variety of estimates in the literature. For this reason, rather than choosing a specific number, we explore how the runtime is affected by different values of the clock rate. The results are shown in Fig.~\ref{fig:time}. For a lower clock rate of 10KHz, the runtime is a few years, even when using only $n_p=4$. On the other hand, assuming an optimistic but in principle achievable clock rate of 100MHz~\cite{tzitrin2021fault,kim2021fault}, for $n_p=4$ we obtain a runtime estimate of less than an a day. For $n_p=9$, the runtime is a about a year. Arguably, an appropriate choice is to select a number of plane waves no larger than $N = 10^6$, which translates to $n_p=7$ as even in a large basis set such as cc-pV5Z basis, our system requires no more than that many planes to be accurate~\cite{su2021fault,babbush2018low}. For this value and a 100MHz clock rate, the runtime is roughly a few weeks.

We reiterate that these are all rough estimates for performing one round of quantum phase estimation. This may need to be repeated to successfully project on the true ground-state. While these estimates require further study, they indicate that additional improvements to all aspects of this algorithm will be crucial for the practicality of quantum algorithms for battery simulation.

\section{Conclusions}
\label{sec:conclusions}
This work presents the first comprehensive analysis of how quantum computers can be used in the context of materials simulation for lithium-ion batteries. In particular, to the best of our knowledge, this is the first attempt to estimate the resources required to execute quantum algorithms aimed at performing high-accuracy ground-state energy calculations of a realistic cathode material. Our study explicitly establishes a connection between battery simulation and quantum computing --- many key properties can be derived from the estimation of ground-state energies of periodic materials, which are amenable to known quantum algorithms. Thus, to impact the field of battery simulations, a focus should be placed on developing quantum algorithms for simulating materials. This includes a closer look at better methods for preparing approximate ground states.

Plane waves are an attractive basis set for describing wave functions of periodic materials as they can inherit the periodicity of the lattice and lead to simpler Hamiltonian representations. In a second-quantized approach where a qubit is assigned to each basis function, this leads to quantum algorithms potentially running on millions of logical qubits, which is a troubling prospect. Consequently, quantum algorithms based on first-quantization could be unmatched for battery simulations that rely heavily on understanding the properties of electrode materials described by periodic systems. First-quantization methods are a relatively new technique in quantum algorithms, which we identify as an important area for future research.

A careful resource estimation of the full quantum algorithm reveals that despite its favorable asymptotic scaling, the overall resource requirements remain daunting. This is true even under the assumption that a Hartree-Fock approximation has sufficiently large overlap wit the true ground state. Concretely, our calculations indicate that thousands of logical qubits and trillions of logical gates are necessary to execute one round of quantum phase estimation. These numbers are not entirely prohibitive; based on optimistic estimates of the clock rate of fault-tolerant quantum computers, implementing the full quantum phase estimation algorithm may take somewhere between hours to months depending on the number of plane waves used. Nevertheless, these resource estimates are a pressing invitation to undertake a dedicated effort aimed at reducing the cost of the quantum algorithm by many orders of magnitude. 

Overall, this manuscript lays the foundation for future work on quantum computing for battery simulation. In the following section, we present an outlook on promising research directions aimed at both increasing the scope of application and improving algorithmic performance.

\section{Outlook}\label{sec:outlook}
The quantum algorithms for battery simulation detailed in this work can be extended to simulate other materials, reduce the gate and qubit costs, and expand the scope of applications by addressing other processes that occur in a battery cell. We provide an outlook on potential avenues for achieving this. 

\subsection{Improving algorithmic performance}

The preparation of the initial state is a crucial step of any quantum phase estimation algorithm. For the simulation of realistic cathode materials, it remains an open question whether the state obtained from a Hartree-Fock approximation has sufficient overlap with the true ground state of the system. It is therefore important to develop techniques to quantify the quality of the input state and to identify better methods for preparing approximate ground states. For example, defining a Slater determinant at the $\Gamma$ point as the initial state is also an approximation that is typically more suitable for simulating a large supercell of the cathode material. Using classical simulations at the level of truncated configuration interaction or coupled-cluster methods for periodic systems could be helpful to quantify the quality of the Hartree-Fock state and, importantly, to build quantum circuits for preparing a better initial state beyond the mean-field approximation. It is also possible to leverage quantum algorithms directly, for example adiabatic quantum algorithms.\\

The gate counts reported in Sec.~\ref{ssec:estimation} show that significant work is likely still required to reduce the computational cost of the quantum algorithm. This could be achieved by reducing both the number of electrons in the computational unit cell and the number of plane-wave basis functions. To this aim, pseudopotentials can be employed to describe the electron-nuclei interaction terms in the Hamiltonian~\cite{martin2020electronic}. This methodology is widely used in DFT approaches to solve problems in materials science. Their inclusion leads to a problem of interacting valence-only electrons where the presence of core electrons is modeled by a short-ranged effective potential, which adds new terms to the electron-nuclei interaction operator $U$~\cite{martin2020electronic}. 

Incorporating the pseudopotentials into the quantum algorithm would require adapting several steps that depend directly on the precise form of the Hamiltonian. This include the decomposition into a linear combination of unitaries and the implementation of the qubitization operator. Further reductions in the cost of running the algorithm may be possible by manipulation of the Hamiltonian, for example by exploiting symmetries or employing factorization strategies. Some of these techniques have already been explored in the context of simulating molecules in second quantization~\cite{lee2021even,motta2021low} and could be extended to deal with periodic materials in first quantization.

\subsection{Extending the scope of application}
As described in Sec.~\ref{sec:quantum_algo}, the algorithm presented in this work is constrained to simulate cathode materials with orthogonal unit cells. Going beyond this approximation is key to simulate different phases of the cathode materials. For example, such a generalization would allow studying state-of-the-art cathodes used in electric vehicles, which crystallize in a rhombohedral structure. Extending the algorithm to account for any crystal system requires generalizing Eqs.~\eqref{eq:dfn_G_p} and~\eqref{eq:pw_2}, which define the reciprocal lattice vectors. While this does not affect the expressions of the Hamiltonian matrix elements, it does require generalizing the decomposition of the Hamiltonian as a linear combination of unitaries, as well as the qubitization operator and its circuit implementation.\\

It is also worth exploring the applicability of the quantum algorithm to simulate battery properties of increasing complexity. For example, the redox potential of the electrolyte molecules is important to predict the electrochemical stability of the cell~\cite{urban2016computational}. Redox potentials are obtained from the ground-state energies of the oxidized and reduced electrolyte molecules, which are embedded in a solvent solution. Single molecules could also be simulated with the present algorithm by using the supercell approach to avoid the interaction between periodic images. In the case of electrolyte molecules, performing accurate simulations of redox potentials requires accounting for solvation effects~\cite{tomasi2005quantum, urban2016computational}. 

More complicated phenomena occur at the electrode/electrolyte interface which are crucial to understand the degradation processes of batteries~\cite{yu2018electrode}. In particular, describing the formation and composition of the solid electrolyte interphase (SEI) is paramount to both improve the performance and to extend the lifespan of lithium-ion batteries~\cite{wang2018review}. Chemical reactions that induce lithium ion losses at the SEI in graphite anodes have been identified as a predominant cause of battery capacity fading upon cycling~\cite{pinson2012theory, barre2013review}. Modeling the growth of the solid electrolyte interphase at the atomic scale is challenging. However, at the core of this process is the reduction of electrolyte molecules near the anode surface~\cite{ramos2016computational}, for which chemical reaction rates can also be computed in terms of ground-state energy calculations. Advanced dynamical simulations of such reactions requires multi-scale approaches combining {\it ab-initio} molecular dynamics and continuum solvent models~\cite{leung2010ab, islam2016reductive, wang2001theoretical}, where most of the computational overhead comes from the costly electronic structure calculations.

It is a new challenge to adapt quantum algorithms to simulate such systems. Typically, quantum algorithms have been studied in the context of molecules and materials consisting of not more than a few hundred electrons. This is largely because, despite their polynomial scaling, quantum algorithms still become more costly when tackling larger systems potentially containing thousands of electrons. This will likely require stepping outside the box of existing approaches and exploring disruptive new ideas in quantum algorithms. Pursuing efforts in this direction may enable feasible simulations of more complicated and larger-scale phenomena that occur in battery cells.\\

As a whole, we are in the early stages of understanding how quantum computing can truly impact industrial operations. This applies in a broad sense to quantum simulation, which extends beyond the context of batteries. Still, lithium-ion batteries are complex systems involving a variety of molecules, materials, and chemical processes. They can therefore serve as testbed for continued development in quantum algorithms, whose gains may then be extended to other areas that benefit from progress in techniques for simulating materials and molecules. Our work is a starting point for the continued developments that will be necessary to understand the role that quantum computers can play in impacting industrial processes, particularly the development of new battery technologies.  

\vspace{0.4cm}

\section{Acknowledgments}
\label{sec:acknowledgments}
The authors thank Tobias J. Osborne, Yuval Sanders, Dominic Berry, Michael Kaicher, Craig Gidney, and Maria Schuld for valuable discussions. P.A.M.C, R.C and M.A.M.-D. acknowledge financial support from the Spanish MINECO grants MINECO/FEDER Projects FIS 2017-91460-EXP, PGC2018-099169-B-I00 FIS-2018 and from CAM/FEDER Project No. S2018/TCS-4342 (QUITEMAD-CM). The research of M.A.M.-D. has been partially supported by the U.S. Army Research Office through Grant No. W911NF-14-1-0103. P.A.M.C. thanks the support of a MECD grant FPU17/03620, and R.C. the support of a CAM grant IND2019/TIC17146. 
\bibliographystyle{apsrev}
\bibliography{references}

\appendix

\section{Basic concepts of periodic systems}
\label{app:ssp}
We summarize the basic concepts used in the paper that are key for the simulation of the electronic structure of periodic materials. More extensive and detailed description can be found in textbooks for solid-state physics and electronic structure methods~\cite{ashcroft1976solid, martin2020electronic}.

\subsection{The direct and reciprocal lattices}
\label{ssec:lattices}
In a crystal structure the positions of the atoms repeat periodically in space. Its entire structure can be defined by specifying (i) the type of atoms and their positions in the smallest portion of the crystal lattice, the primitive unit cell, and (ii) the primitive vectors $\bm{a}_1, \bm{a}_2, \bm{a}_3$ used to define all possible translations in space. The lattice of points obtained by replicating the unit cell is called the {\it Bravais} (direct) lattice. For a three-dimensional space the direct lattice consists of all points with positions vectors
\begin{equation}
\bm{R}_n = n_1 \bm{a}_1 + n_2 \bm{a}_2 + n_3 \bm{a}_3,
\label{eq:d_lattice}
\end{equation}
where $n_1$, $n_2$ and $n_3$ take integer values. For example, for the simple case of an orthogonal lattice, the primitive vectors are given by
\begin{equation}
\bm{a}_1 = a_1 \hat{\bm{x}}, \quad \bm{a}_2 = a_2 \hat{\bm{y}}, \quad \bm{a}_3 = a_3 \hat{\bm{z}},
\label{eq:orth_lattice}
\end{equation}
where $a_1$, $a_2$ and $a_3$ are the lattice constants defining the distance between the atoms in different unit cells along the orthogonal directions. More complicated primitive vectors to describe different types of materials are extensively covered in the literature~\cite{ashcroft1976solid, prince2004international}.

The {\it primitive} unit cell defines a volume that fills all the space without leaving gaps when it is translated through all the vectors in a Bravais lattice. The {\it conventional} unit cell fills the same space when translated through some subset of the vectors of the lattice. It is typically larger than the primitive cell and contains the crystal symmetry. The primitive cell with the full symmetry of the lattice is known as the {\it Wigner-Seitz} cell, which is defined by the space bounded by the planes that bisect the lines joining one site of the lattice with all its closest neighbors~\cite{ashcroft1976solid}.

On the other hand, the concept of a reciprocal lattice is fundamental for both analytical and numerical techniques to simulate periodic systems. Consider a set of points $\bm{R}$ constituting a direct lattice and a plane wave $e^{i\bm{k} \cdot \bm{r}}$. The set of all wave vectors $\bm{G}$ that yield plane waves with the periodicity of a given direct lattice is known as its reciprocal lattice. This periodicity restriction implies that the condition
\begin{equation}
e^{i{\bm{G} \cdot (\bm{r} + \bm{R})}} = e^{i \bm{G} \cdot \bm{r}} \implies \quad e^{i{\bm{G} \cdot \bm{R}}} = 1,
\label{eq:r_latt_1}
\end{equation}
applies for any $\bm{r}$ and for all $\bm{R}$ in the direct lattice. For a given set of primitive vectors $\bm{a}_1$, $\bm{a}_2$, $\bm{a}_3$, the reciprocal lattice can be generated by the primitive vectors
\begin{eqnarray}
&&\bm{b_1} = 2\pi \frac{\bm{a}_2 \times \bm{a}_3}{\bm{a}_1 \cdot (\bm{a}_2 \times \bm{a}_3)} \nonumber\\
&&\bm{b}_2 = 2\pi \frac{\bm{a}_3 \times \bm{a}_1}{\bm{a}_1 \cdot (\bm{a}_2 \times \bm{a}_3)} \nonumber \\
&&\bm{b}_3 = 2\pi \frac{\bm{a}_1 \times \bm{a}_2}{\bm{a}_1 \cdot (\bm{a}_2 \times \bm{a}_3)},
\end{eqnarray}
which satisfy $\bm{b}_i \cdot \bm{a}_j = 2\pi \delta_{ij}$ with $i, j = 1, 2, 3$. For example, from Eq.~\eqref{eq:orth_lattice} it follows that the primitive vectors $\bm{b}_i$ for an orthogonal lattice are defined as
\begin{equation}
\bm{b}_1 = \frac{2\pi}{a_1} \hat{\bm{x}}, \quad \bm{b}_2 = \frac{2\pi}{a_2} \hat{\bm{y}}, \quad \bm{b}_3 = \frac{2\pi}{a_3} \hat{\bm{z}}.
\label{eq:rec_orth_lattice}
\end{equation}
Using Eq.~\eqref{eq:r_latt_1} it can be shown that the reciprocal lattice associated with a given direct lattice consists of all points with position vectors 
\begin{equation}
\bm{G}_n = n_1 \bm{b}_1 + n_2 \bm{b}_2 + n_3 \bm{b}_3.
\label{eq:r_lattice}
\end{equation}
The Wigner-Seitz primitive cell of the reciprocal lattice is called the first {\it Brillouin} zone.

\subsection{Single-electron states in a periodic potential}
\label{ssec:bands}
In the independent electron approximation the effective potential felt by an electron in a crystal structure has the periodicity of the underlying Bravais lattice:
\begin{equation}
U^\mathrm{eff}(\bm{r} + \bm{R}) = U^\mathrm{eff}(\bm{r}).
\label{eq:crystal_pot}
\end{equation}
It follows from Bloch's theorem that the wave function of a single electron in the periodic potential $U^\mathrm{eff}(\bm{ r})$ can be chosen to have the form
\begin{equation}
\phi_{\bm{k}} (\bm{r}) = e^{i \bm{k} \cdot \bm{r}} u_{\bm{k}}(\bm{ r}),
\label{eq:bloch_wf}
\end{equation}
where $u_{\bm{k}}(\bm{r})$ has the periodicity of the Bravais lattice. By imposing the Born-Von Karman boundary condition on the wave function, it is straightforward to show that the allowed values of $\bm{k}$, known as k-points, are given by the expression
\begin{equation}
\bm{k} = \sum_{i=1}^3 \frac{n_i}{N_i} \bm{b}_i,
\label{eq:k_points}
\end{equation}
where $N_i$ are integers of order $N_\mathrm{cell}^{1/3}$ and $N_\mathrm{cell}=N_1N_2N_3$ is the total number of unit cells in the crystal. In the limit of the macroscopic crystal, $\bm{k}$ can be considered a continuous variable which takes values in the first Brillouin zone of the reciprocal lattice.

In general, a wave function $\phi(\bm{r})$ that satisfies the Schr\"odinger equation
\begin{equation}
H^\mathrm{eff} \phi(r) = \left[ -\frac{\nabla^2}{2} + U^\mathrm{eff}(\bm{ r}) \right] \phi(r) = E \phi(r),
\label{eq:sch_1e}
\end{equation}
can be expanded in a set of plane waves that satisfy the boundary conditions:

\begin{equation}
\phi(\bm{r}) = \sum_{\bm{q}} C_{\bm{q}} e^{i \bm{q} \cdot \bm{r}}.
\label{eq:expansion}
\end{equation}
Similarly, we can expand the effective potential $U^\mathrm{eff}(\bm{r})$ using a set of plane waves. Since  $U^\mathrm{eff}(\bm{r})$ is periodic in the lattice, its expansion will only contain plane waves with wave vectors that are vectors of the reciprocal lattice

\begin{eqnarray}
&&U^\mathrm{eff}(\bm{r}) = \sum_\mu U^\mathrm{eff}(\bm{G}_\mu) e^{i \bm{G}_\mu \cdot \bm{r}} \label{eq:u_exp}, \\
&& U^\mathrm{eff}(\bm{G}) = \frac{1}{\Omega} \int_\mathrm{cell} d\bm{ r} U^\mathrm{eff}(\bm{r}) e^{-i \bm{G} \cdot \bm{r}},
\end{eqnarray}
where $\Omega$ denotes the volume of the unit cell. Next, we use Eqs.~\eqref{eq:expansion} and \eqref{eq:u_exp} to represent the Schr\"odinger equation in the basis of plane waves and define $\bm{q} = \bm{k} + \bm{G}_\mu$ to obtain the equation for the coefficients $C$ representing the single-electron states $\phi_n(\bm{r})$ in the plane wave basis, Eq.~\eqref{eq:expansion} \cite{martin2020electronic}
\begin{equation}
\sum_{\mu^\prime} H^{\mathrm{eff}}_{\mu \mu^\prime}(\bm{k}) ~ C_{\mu^\prime n}(\bm{ k}) = E_n (\bm{k}) C_{\mu n}(\bm{k}),
\label{eq:coeffs}
\end{equation}
where the Hamiltonian matrix is defined as

\begin{equation}
H^\mathrm{eff}_{\mu \mu^\prime}(\bm{k}) = \frac{\vert\vert \bm{k} + \bm{G}_\mu \vert\vert^2}{2} \delta_{\mu\mu^\prime} + U^\mathrm{eff}(\bm{G}_\mu - \bm{G}_{\mu^\prime}).
\end{equation}

Summarizing:

\begin{enumerate}
    \item Eq.~\eqref{eq:coeffs} is the Schr\"odinger equation in momentum space, simplified by the fact that $U^\mathrm{eff}(\bm{k})$ is nonvanishing only when $\bm{k}$ is a vector of the reciprocal lattice.
    
    \item For a fixed $\bm{k}$, the set of equations for all reciprocal lattice vectors $\bm{G}$ couple only those coefficients $C_{\bm{k}}$, $C_{\bm{k} + \bm{G}_1}$, $C_{\bm{k} + \bm{G}_2}$, $\dots$ whose wave vectors differ from $\bm{k}$ by a reciprocal lattice vector.
    
    \item The eigenvalues $E_n(\bm{k})$ and the eigenvectors $C_{\mu n}(\bm{k})$ are characterized by the discrete band index $n$.
    
    \item The number of bands for each k-point is determined by the number of plane waves entering the expansion~\eqref{eq:u_exp}.
    
    \item In practice, the plane wave basis is truncated using a cutoff value for the kinetic energy:
    \begin{equation}
    \frac{\vert\vert \bm{k} + \bm{G}_\mu\vert\vert^2 }{2} < E_\mathrm{cutoff}.
    \label{eq:cutoff}
    \end{equation}
    
    \item The wave function $\phi_{n\bm{k}}(\bm{r})$ is a superposition of plane waves of the form:
    \begin{equation}
      \phi_{n\bm{k}}(\bm{r}) = \frac{1}{\sqrt{N_\mathrm{cell}\Omega}}\sum_\mu C_{\mu n}(\bm{k})  e^{i (\bm{k} + \bm{G}_\mu) \cdot {\bf r}}.
    \end{equation}
\end{enumerate}

\begin{figure*}
\mbox{
\Qcircuit @C=0.9em @R=0.8em {
& {/}\qw & \gate{\text{Exponential}} & \qw  & \ket{\mu} & & \qw  & \ctrl{1} & \ctrl{2} & \ctrl{3} & \qw& \qw& \qw & \qw & \ctrl{1} & \ctrl{1} & \qw\\
& {/}\qw & \qw & \qw & \qw & \qw & \qw & \gate{H} &\qw & \qw & \qw & \ket{\nu_x} & & \ctrl{1}& \ctrl{1}& \ctrl{1} & \qw \\
& {/}\qw & \qw & \qw & \qw & \qw & \qw & \qw & \gate{H} & \qw & \qw & \ket{\nu_y} & & \ctrl{1}& \ctrl{1}& \ctrl{1} & \qw \\
& {/}\qw & \qw & \qw & \qw & \qw & \qw & \qw & \qw & \gate{H} & \qw & \ket{\nu_z} & & \ctrl{2}& \ctrl{3}& \ctrl{1} & \qw \\
& {/}\qw & \gate{H} & \qw & \ket{m} &  & \qw & \qw & \qw & \qw & \qw & \qw & \qw & \qw & \qw & \ctrl{3} & \qw  \\
&\ket{0}_{\text{flag}} && \qw & \qw & \qw & \qw & \qw & \qw & \qw & \qw & \qw & \qw & \gate{\bm{\nu}\neq -0} & \qw & \qw & \qw \\
&\ket{0}_{\text{flag}} && \qw & \qw & \qw & \qw & \qw & \qw & \qw & \qw & \qw & \qw & \qw & \gate{\bm{\nu}\in B_\mu} & \qw & \qw \\
&\ket{0}_{\text{flag}} && \qw & \qw & \qw & \qw & \qw & \qw & \qw & \qw & \qw & \qw & \qw & \qw & \gate{(2^{\mu-2})^2 M > m \|\bm{\nu}\|^2} & \qw \\
}
}
\caption{\textbf{Quantum circuit for momentum state preparation}. The circuit for implementing a state with exponential amplitudes is the same as in Fig.~\ref{fig:prep_r} in the main text. The controlled Hadamard gates correspond to the preparation of $\mathcal{C}_\mu$ in Fig.~\ref{fig:prep_C_mu}. There is also a register for the uniform superposition over $\ket{m}$, as well as the three tests. The later one checking the condition $(2^{\mu-2})^2 M > m \|\bm{\nu}\|^2$ constitutes the key step in this procedure.}
\label{fig:prep_momentum}
\end{figure*}
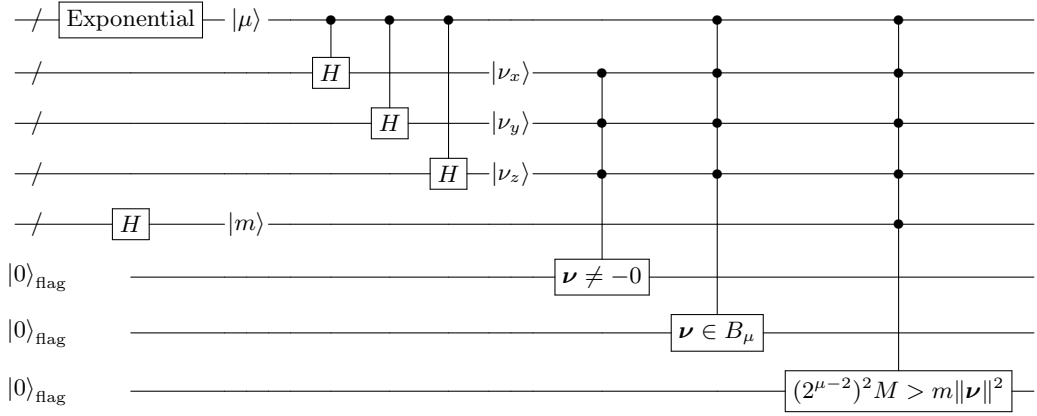

\section{Electronic Hamiltonian in first quantization}
\label{app:hamiltonian}

We derive the matrix elements of the Hamiltonian in first quantization given by Eqs.~\eqref{eq:t_op}-\eqref{eq:v_op} in Sec.~\ref{ssec:hamiltonian}. In first quantization, each term of the electronic Hamiltonian $H=T+U+V$ can be projected on to the plane-wave basis functions defined in Eq.~\eqref{eq:pw_1} as follows:
\begin{eqnarray}
&& T = \sum_{i=1}^\eta \sum_{p,q=1}^N T_{pq} \vert \bm{p} \rangle \langle \bm{q} \vert_i \label{eq:app_1},\\
&&U = \sum_{i=1}^\eta \sum_{p,q=1}^N U_{pq} \vert \bm{p} \rangle \langle \bm{q} \vert_i \label{eq:app_2},\\
&& V = \frac{1}{2}\sum_{i \neq j = 1}^{\eta_\mathrm{val}} \sum_{p, q, r, s=1}^N V_{pqrs} \vert \bm{p} \rangle \langle \bm{s} \vert_i ~ \vert \bm{q} \rangle \langle \bm{r} \vert_j \label{eq:app_3},
\end{eqnarray}
with $\ket{\bm{p}}_i$ and $\ket{\bm{q}}_i$ indexing momentum basis functions. The matrix elements of the kinetic energy operator are obtained from the integral
\begin{eqnarray}
T_{pq} &=& \int d\bm{r} \varphi_p^*(\bm{r}) \bigg(-\frac{\nabla^2}{2} \bigg) \varphi_q(\bm{r}) \nonumber \\
&=& \frac{\| \bm{G}_q \|^2}{2\Omega} \int d\bm{r} e^{i(\bm{G}_q - \bm{G}_p)} = \delta_{p, q} \frac{\vert\vert \bm{G}_p \vert\vert^2}{2}.
\label{eq:t_pq}
\end{eqnarray}
For computing the matrix elements of the electron-nuclei and electron-electron interaction, we use that the Fourier transform of the Coulomb potential $1/r$ is $\mathcal{F}[1/r] = 4\pi/\bm{G}^2$. Additionally, we employ Eq.~\eqref{eq:h_en} to compute the matrix elements of the one-particle operator $U$ as
\begin{eqnarray}
\kern+25pt &U_{pq}&=\sum_{I=1}^L \int d\bm{r} \varphi_p^*(\bm{r}) \bigg(-\frac{Z_I}{\vert\vert \bm{r}-\bm{R}_I \vert\vert}\bigg) \varphi_q(\bm{r}) \nonumber \\
&=& -\frac{1}{\Omega}\sum_{I=1}^L Z_I \int d\bm{r} \frac{e^{-i(\bm{G}_p - \bm{G}_q) \cdot (\bm{r} - \bm{R}_I)}}{\vert\vert \bm{r}-\bm{R}_I \vert\vert} e^{i(\bm{G}_q - \bm{G}_p) \cdot \bm{R}_I} \nonumber \\
&=& -\frac{4\pi}{\Omega} \sum_{I=1}^L Z_I \frac{e^{i(\bm{G}_q - \bm{G}_p) \cdot \bm{R}_I}}{\vert\vert \bm{G}_p - \bm{G}_q \vert\vert^2}.
\label{eq:u_pq}
\end{eqnarray}
Similarly, for the matrix elements of the two-particle operator $V$ we have
\begin{eqnarray}
\kern+25pt &V_{pqrs}& = \int d\bm{r}_1 d\bm{r}_2 \frac{\varphi_p^*(\bm{r}_1) \varphi_q^*(\bm{r}_2) \varphi_r(\bm{r}_2) \varphi_q(\bm{r}_1)}{\vert\vert \bm{r}_1 - \bm{r}_2 \vert\vert} \nonumber \\
&=& \frac{4\pi}{\Omega^2 \vert\vert \bm{G}_p - \bm{G}_s \vert\vert^2} \int d\bm{r}_2 e^{i[(\bm{G}_r - \bm{G}_q) - (\bm{G}_p - \bm{G}_s)]  \cdot \bm{r}_2} \nonumber \\
&=& \frac{4\pi}{\Omega} \frac{\delta_{\bm{G}_p - \bm{G}_s, \bm{G}_r - \bm{G}_q}}{\vert\vert \bm{G}_\nu \vert\vert^2},
\label{eq:v_pqrs}
\end{eqnarray}
where $\bm{G}_\nu = \bm{G}_p - \bm{G}_s = \bm{G}_r - \bm{G}_q \neq 0$. By inserting the Eqs.~\eqref{eq:t_pq}-\eqref{eq:v_pqrs} into the Eqs.~\eqref{eq:app_1}-\eqref{eq:app_3}, correspondingly, we obtain the Hamiltonian representation given by Eqs.~\eqref{eq:t_op}-\eqref{eq:v_op}. Note that in those equations, we require that $\nu \in \mathcal{G}_0$. This aliasing is commonplace in electronic structure codes, and the error caused has the same asymptotic behavior as the basis error~\cite{babbush2018low}.

\section{Preparation of the momentum state}\label{app:Momentum_state}
\begin{figure}
        \mbox{
        \Qcircuit @C=0.9em @R=0.75em { 
          &  &  & {\gategroup{1}{4}{5}{4}{.7em}{\{}} & \qw & \ctrl{8} & \qw & \qw& \qw & \qw& \qw\\
          & & & & \qw & \qw & \ctrl{8} & \qw & \qw & \qw& \qw\\
          & \ket{\mu}& & &\vdots & & & & \\
          & & & & & & & & \\
          & & & & \qw & \qw & \qw & \ctrl{8}& \qw& \qw& \qw\\
          & & & & & & & & \\
          & & & & & & & & \\
          &  &  & {\gategroup{8}{4}{13}{4}{.7em}{\{}} & \gate{H} & \qw & \qw & \qw& \qw & \ctrlo{1}& \qw\\
          &  &  &  & \qw & \gate{H} & \qw & \qw& \qw & \ctrlo{1}& \qw\\
          & \ket{\nu_i} & & & \qw & \qw & \gate{H} & \qw  & \qw & \ctrlo{0} & \qw\\
          & & & &\vdots & & & & & \vdots\\
          & & & & & & & & \\
          & & & & \qw &  \qw & \qw & \gate{H} & \qw & \ctrlo{1}& \qw\\
          & \ket{0}_{\text{flag}} & & &  \qw &  \qw& \qw &  \qw & \qw & \targ & \qw
        }
        }
        \caption{\textbf{Preparation of the superposition corresponding to $\mathcal{C}_\mu$}. The first register in $\ket{\nu_i}$ is the sign qubit, using controlled Hadamard gates. This procedure has to be repeated for $i\in\{x,y,z\}$. The last multi-controlled not can be understood as part of the detection of $\bm{\nu}$ having a $-0$ value in one of the components.}
        \label{fig:prep_C_mu}
\end{figure}
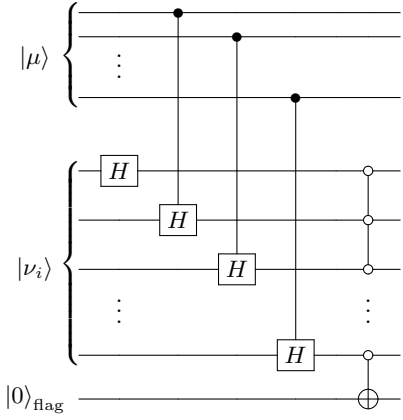

The process of implementing PREP$_U$ and PREP$_V$ involves the common step of preparing the momentum state~\eqref{eq:1/nu_state} reproduced below: $$\frac{1}{\sqrt{\lambda_{\nu}}}\sum_{\nu \in \mathcal{G}_0} \frac{1}{\|\bm{\nu}\|}\ket{\nu_x}\ket{\nu_y}\ket{\nu_z}.$$
To prepare it~\cite[Pag. 4-5]{babbush2019quantum}, the necessary steps as depicted in Fig.~\ref{fig:prep_momentum} are:
\begin{enumerate}
    \item Use the same technique as in Fig.~\ref{fig:prep_r} for $\text{PREP}_T$ to prepare a unary-encoded register
    \begin{equation}\label{eq:PREP_U+V_mu_state}
    \frac{1}{\sqrt{2^{n_p+2}}}\sum_{\mu= 2}^{n_p+1}\sqrt{2^{\mu}}\ket{\mu} = \frac{1}{\sqrt{2^{n_p+2}}}\sum_{\mu= 2}^{n_p+1}\sqrt{2^{\mu}}\ket{0\ldots\underbrace{1\ldots1}_{\mu}}.
    \end{equation}
    
    \item Prepare a uniform superposition state using controlled Hadamards over registers $\ket{\nu_x}$, $\ket{\nu_y}$, and $\ket{\nu_z}$, which will take values from $-2^{\mu-1}+1$ to $2^{\mu-1}-1$ as signed integers. These superpositions can be written using a series of nested cubes $\mathcal{C}_\mu$ and their differences $B_\mu = \mathcal{C}_\mu \backslash \mathcal{C}_{\mu - 1}$; see the circuit depicted in Fig.~\ref{fig:prep_C_mu}.
    
    \item The previous preparation contains both a representation for $\ket{+0}$ and $\ket{-0}$. The latter is therefore flagged as failure.
    
    \item Similarly, to avoid double-counting, we should flag as failure when $\bm{\nu}$, prepared for a given value of $\mu$, is also in the inner cube $\mathcal{C}_{\mu -1}$, i.e. $\bm{\nu} \notin B_\mu$.
    
    \item Use Hadamard gates to prepare a superposition over $\ket{m}$ from $0$ to $M$, where $M$ is a large power of two.
    
    \item Finally, this last register undergoes an inequality test
    \begin{equation}\label{eq:comparison_PREP_U+V}
        (2^{\mu-2})^2 M > m \|\bm{\nu}\|^2.
    \end{equation}
    This test \cite[Eq.~(84)]{su2021fault} yields
    \begin{equation}\label{eq:PREP_U+V_result_state}
    \begin{split}
      \frac{1}{\sqrt{M(2^{n_p+2})}}\sum_{\mu=2}^{n_p+1}\sum_{\bm{\nu}\in B_{\mu}}\sum_{m=0}^{\lceil M(2^{\mu-2}/\|\bm{\nu}\|)^2\rceil-1}\\\times \frac{1}{2^\mu}\ket{\mu}\ket{\nu_x}\ket{\nu_y}\ket{\nu_z}\ket{m}\ket{0} + \ket{\Psi^\perp},
      \end{split}
    \end{equation}
    with the desired amplitudes for each $\bm{\nu}$ upon success:
    \begin{equation}
        \sqrt{\frac{\lceil M(2^{\mu-2}/\|\bm{\nu}\|)^2\rceil}{M2^{2\mu}(2^{n_p+2})}}\approx \frac{1}{4\sqrt{2^{n_p+2}}}\frac{1}{\|\bm{\nu}\|}.
    \end{equation}
    Note the amplitudes $\frac{1}{2^\mu}$ in \eqref{eq:PREP_U+V_result_state} come from the factor $\sqrt{2^{\mu}}$ in \eqref{eq:PREP_U+V_mu_state}, as well as three factors of $\sqrt{2^{-\mu}}$ from the uniform superposition over $\ket{\nu_x}$, $\ket{\nu_y}$ and $\ket{\nu_z}$.
\end{enumerate}

\section{Implementing SEL$_U$ and SEL$_V$ operators}\label{app:sel_operators}

We explain in more depth the steps to implement SEL$_U$ and SEL$_V$, implied from \eqref{eq:U_operator} and \eqref{eq:V_operator}: (i) controlled sums and subtractions, (ii) a phase to cancel out the amplitudes of the invalid states, and (iii) exclusively for SEL$_U$, the phase $-e^{i\bm{G}_{\nu}\cdot \bm{R}_I}$.

Following the procedure depicted in Fig.~\ref{fig:SEL_technique}, sums and subtractions are performed by the operator $O$. The details and cost involved in this arithmetic operation can be found in \cite[Sec.~II.D]{su2021fault}.

The control-phase cancellations $(-1)^{b[(\bm{q}-\bm{\nu})\notin \mathcal{G}]}$ and $(-1)^{b([\bm{p}+\bm{\nu}\notin \mathcal{G}]\vee[\bm{q}-\bm{\nu}\notin \mathcal{G}])}$ are similar to the case discussed for $\text{SEL}_T$. For example, if $\bm{p+\nu}\notin \mathcal{G}$, then one of the three coordinates of  $\bm{p+\nu}$ has absolute value larger than $2^{n_p-1}$, which means that some extra auxiliary qubit will take value $\ket{1}$. This qubit can be used to apply a multi-controlled $Z$ gate on $\ket{+}_b$. In a further optimization, this last phase implementation can be shown to be unnecessary when $\bm{p+\nu}$ or $\bm{q-\nu}$ are outside $\mathcal{G}$, since the extra auxiliary qubits are among those that are automatically selected to be $\ket{0}$ in the block-encoding identity~\eqref{eq:block-encoding_id}~\cite[Sec.~II.D]{su2021fault}. Notice how this selection also enables the implementation of the non-unitary operators $H_{\ell_U}, H_{\ell_V}$, since any state $\bm{p+\nu}$ or $\bm{q-\nu}$ outside of $\mathcal{G}$ is projected to zero, which is how $H_{\ell_U}, H_{\ell_V}$ act on these states.

The phase $-e^{i\bm{G}_{\nu}\cdot \bm{R}_I}$ in SEL$_U$ requires to multiply and sum all three coordinates $G_{\nu_i} (\bm{R}_I)_i$. The inner product can be done in the computational basis with standard reversible quantum algorithms. Finally, the binary expression of $\sum_i G_{\nu_i} (\bm{R}_I)_i$ is used to perform controlled $R_Z(\pi/2^{b+1})$ rotations, where $b$ is the bit we are rotating.

Overall, we may describe the SEL$_U$ operator as
\begin{equation}
\begin{split}
    \text{SEL}_U: \ket{b}_b\ket{j}_e\ket{0}_m\ket{\bm{\nu}}_k\ket{\bm{R}_I}_l\ket{\bm{q}_j}\mapsto \\
    \ket{b}_b\ket{j}_e\ket{0}_m\ket{\bm{\nu}}_k\ket{\bm{R}_I}_l\ket{\bm{q}_j-\bm{\nu}}\mapsto\\
    (-1)^{b[(\bm{q}-\bm{\nu})\notin \mathcal{G}]}\ket{b}_b\ket{j}_e\ket{0}_m\ket{\bm{\nu}}_k\ket{\bm{R}_I}_l\ket{\bm{q}_j-\bm{\nu}}\mapsto\\
    -e^{i\bm{G}_{\nu}\cdot \bm{R}_I}(-1)^{b[(\bm{q}-\bm{\nu})\notin \mathcal{G}]}\ket{b}_b\ket{j}_e\ket{0}_m\ket{\bm{\nu}}_k\ket{\bm{R}_I}_l\ket{\bm{q}_j-\bm{\nu}}.
\end{split}
\end{equation}

Similarly, for SEL$_V$ we apply the transformation 
\begin{equation}
\begin{split}
    \text{SEL}_V: \ket{b}_b\ket{i}_d\ket{j}_e\ket{1}_m\ket{\bm{\nu}}_k\ket{\bm{p}_i}\ket{\bm{q}_j}\mapsto \\
    \ket{b}_b\ket{i}_d\ket{j}_e\ket{1}_m\ket{\bm{\nu}}_k\ket{\bm{p}_i+\bm{\nu}}\ket{\bm{q}_j-\bm{\nu}}\mapsto\\
    (-1)^{b([\bm{p}+\bm{\nu}\notin \mathcal{G}]\vee[\bm{q}-\bm{\nu}\notin \mathcal{G}])}\ket{b}_b\ket{i}_d\ket{j}_e\ket{1}_m\ket{\bm{\nu}}_k\\
    \otimes\ket{\bm{p}_i+\bm{\nu}}\ket{\bm{q}_j-\bm{\nu}}.
\end{split}
\end{equation}
It can be seen from these two equations that the operation $\ket{\bm{q}_j}\ket{\bm{\nu}}\mapsto \ket{\bm{q}_j-\bm{\nu}}\ket{\bm{\nu}}$ must be implemented in both cases, so it can be implemented just once controlled on the register that selects $U + V$ instead of $T$.

\section{Toffoli gate cost full equation}\label{app:toffoli_cost}

As mentioned in Sec.~\ref{sssec:Gate_cost}, below we reproduce the full Toffoli gate cost equation of the qubitization-based quantum phase estimation algorithm, while briefly outlining the origin of each term in the expression:

\begin{equation}\label{eq:cost}
\begin{split}
\underbrace{\left\lceil\frac{\pi \lambda}{2\varepsilon_{\text{QPE}}}\right\rceil}_{\text{\#(controlled-Q calls)}}\Big(  \underbrace{2(n_T + 4n_{\eta Z}+2b_r -12)}_{\text{preparation qubit } T/(U+V)} + \\
 +\underbrace{14n_{\eta} + 8b_r - 36}_{\text{uniform }i \& j \text{ and $i\neq j$ test}}\\
+ \underbrace{a[3n^2_p+15n_p-7+4n_M(n_p+1)]}_{\text{preparation } \bm{1/|\nu|} \text{ amplitudes}}
+\underbrace{\lambda_Z + Er(\lambda_Z)}_{\text{QROM}} \\
+ \underbrace{2(2n_p + 2b_r - 7)}_{\text{preparation over }w,r \&s} + \underbrace{12\eta n_p}_{\text{swap } p \&q} +  \underbrace{5(n_p-1)+2}_{\text{SEL}_T}\\
+ \underbrace{24n_p}_{\ket{\bm{p}\pm \bm{\nu}}} + \underbrace{6n_pn_R}_{e^{i\bm{G}_{\nu}\cdot \bm{R}_I}}+\underbrace{18}_{\text{selection between } T,U,V}+\\
\underbrace{n_{\eta Z} + 2n_\eta + 6n_p + n_M + 16}_{(2\ket{0}\bra{0} - \bm{1})}+ \underbrace{\Tilde{O}(\log\varepsilon^{-1})}_{\text{Rotations}}\Big),
\end{split}
\end{equation}

The different terms of the form $n_x$ as well as the term $b_r$ denote qubits numbers. Importantly, all these quantities are logarithmic in the precision derived from various error sources. In addition, note that $a=3$ or $1$ depending on whether or not amplitude amplification is used in the preparation of $1/\|\bm{\nu}\|$ amplitudes (see part 2 in Sec.~\ref{ssec:Circuit_Implementation}).

\section{Non-cubic unit cells}\label{app:non_cubic}

We explore what happens if the cell, instead of having a cubic form, is a rectangular parallelepiped, i.e., the primitive vectors of the cell are orthogonal but not orthonormal. Therefore, let us take the vectors of the direct lattice~\eqref{eq:d_lattice} to be $\left(a_1 n_1, a_2n_2, a_3n_3\right)$, where the coefficients $a_i$ are different.

Following the App.~\ref{app:hamiltonian} formalism to find the Hamiltonian matrix elements, we observe that the Fourier transform of the Coulomb potential $1/r$ still has the same form, i.e. $\mathcal{F}[1/r] = \frac{4\pi}{||\bm{G}_\nu||^2}$ . Thus, when expressing the Hamiltonian operators in Eqs.~\eqref{eq:t_op},~\eqref{eq:u_op} and~\eqref{eq:v_op}, the components of $\bm{G}_{\nu}$~\eqref{eq:dfn_G_p} are rescaled appropriately. Notice that $\bm{p}$ or $\bm{\nu}$ do not change as they label the plane wave basis along with only the vector space structure~\eqref{eq:pw_2}, while the geometry is accounted for in $\bm{G}_{\nu}$~\eqref{eq:dfn_G_p}. This has the following consequences for the algorithm:
\begin{itemize}
    \item In $\text{PREP}_T$, we previously created a uniform superposition over $w$, with $w$ indexing each component of $\bm{\nu}$. Now such a superposition will not be done uniformly, but according to the weights $1/a_i$.
    \item In $\text{PREP}_{U+V}$, to prepare the momentum state $\sum_{\bm{\nu}}\frac{1}{\|\bm{G}_{\nu}\|} \ket{\bm{\nu}}$, each $\|\bm{\nu}\|$ in the equations of App.~\ref{app:Momentum_state} needs to be replaced with $\|\bm{G}_{\nu}\|$. Thus, there is a rescaling of the amplitudes in~\eqref{eq:comparison_PREP_U+V}, i.e. $(2^{\mu-2})^2 M > m (\nu_x^2/a_1^2+\nu_y^2/a_2^2+\nu_z^2/a_3^2)$. The remaining inequality test can be carried out similarly yielding the desired amplitudes. Notice that we normalize the coefficients $a_i^{-1}$. For our case-study constants and after amplitude amplification, this has the effect of increasing the asymptotic failure probability from $\sim$0.1\% to $\sim$5.5\%.
\end{itemize}
Regarding the $\text{SEL}$ operators, the only phase which could have a change in its implementation is $-e^{i\bm{G}_{\nu}\cdot \bm{R}_I}$. However, since the $\bm{R}_I$ coordinates are those of the direct lattice with coordinates $(R_I)_i a_i$, which have the inverse weights of $\bm{G}_\nu$'s coordinates $p_i/a_i$, there is no change to the implementation of this phase either.
\end{document}